\documentclass[a4paper,11pt]{article}

\usepackage{jheppub} 
\usepackage[T1]{fontenc} 
\usepackage[textsize=footnotesize,textwidth=2.5cm]{todo notes}

\makeatletter

\newcommand{\Rmnum}[1]{\expandafter\@slowromancap\romannumeral #1@}

\makeatother
\definecolor{mikadoyellow}{rgb} {0.16, 0.254, 0.6}
\usepackage{amssymb,amsmath,latexsym,bm,amsfonts}
\usepackage{graphicx}
\usepackage{longtable}
\usepackage{color,xcolor}
\usepackage{indentfirst}
\usepackage{float}
\usepackage{subfigure}
\usepackage{comment}
\usepackage[normalem]{ulem}
\usepackage{verbatim}
\usepackage{tikz}
\usepackage{braket}
\usepackage[nameinlink]{cleveref}
\newcommand{\vast}{\bBigg@{3}}
\newcommand{\Vast}{\bBigg@{5}}
\newcommand{\TTbar}{\text{T}\bar{\text{T}}}
\newcommand{\ttbar}{T\bar{{T}}}
\newcommand{\zbar}{\raisebox{0.2ex}{--}\kern-0.6em Z}

\def\CD{{\cal D}}

\def\CH{{\cal H}}

\def\CL{{\cal L}}
\def\CR{{\cal R}}
\def\CM{{\cal M}}

\def\CO{{\cal O}}
\def\CP{{\cal P}}

\def\CS{{\cal S}}

\def\CZ{{\cal Z}}
\def\BC{\mathbb{C}}

\def\BR{\mathbb{R}}
\def\BS{\mathbb{S}}

\def\BZ{\mathbb{Z}}

\def\d{\textrm{d}}
\def\del{\partial}
\def\ttbar{\textrm{T}\bar{\textrm{T}}}
\def\TTbar{\textrm{T}\overline{\textrm{T}}}

\title{The holographic $\textrm{T}\overline{\textrm{T}}$ deformation of the CFT$_2$ with gravitational anomalies}

\author[a]{Debarshi Basu, Qiang Wen, Mingshuai Xu}

\affiliation[a]{Shing-Tung Yau Center and School of Physics, Southeast University, Nanjing 210096, China}

\emailAdd{debarshi.128@gmail.com, wenqiang@seu.edu.cn, xumingshuai@seu.edu.cn}

\abstract{We develop the holographic framework for the $\TTbar$ deformation of two-dimensional conformal field theories (CFT$_2$) with gravitational anomalies, characterized by unequal left and right central charges and holographically dual to topological massive gravity (TMG). Utilizing the mixed boundary condition prescription, we construct the deformed BTZ black hole geometry and derive the corresponding deformed energy spectrum, confirming that the universal flow equation remains valid despite the presence of gravitational anomalies. From the boundary perspective, we compute leading-order corrections to entanglement entropy and reflected entropy induced by the $\TTbar$ deformation, as well as the balanced partial entanglement entropy non-perturbatively. On the gravity side, these quantities are evaluated using spinning worldlines in the deformed bulk geometry, with results matching their field-theoretic counterparts in the high-temperature limit. We further analyze the reality condition for holographic entanglement entropy, which constrains the deformation parameter and reveals a generalized Hagedorn behavior. This Hagedorn-like transition is also independently reproduced from the asymptotic density of states in the deformed anomalous CFT$_2$, providing additional evidence for its universality.
	}

\begin{document} 
	\maketitle
	\flushbottom
	\section{Introduction}
	
	Two-dimensional conformal field theories (CFT$_2$) have long served as a cornerstone of theoretical physics, offering deep insights into quantum field theory, statistical mechanics, and string theory. Their exact solvability, infinite-dimensional symmetry algebra, and modular properties render them ideal laboratories for studying non-perturbative dynamics and holography. In particular, the AdS$_3$/CFT$_2$ correspondence \cite{Brown:1986nw,Strominger:1997eq,Maldacena:1997re} provides a compelling realization of the holographic principle, where the bulk geometry is reconstructed from boundary CFT data.
	
	A major advancement in this context is the discovery of the solvable yet \emph{irrelevant} $\TTbar$ deformation of two-dimensional quantum field theories \cite{Zamolodchikov:2004ce,Smirnov:2016lqw,Cavaglia:2016oda}. This deformation, generated by a bilinear composite operator constructed from the components of the stress-energy tensor,
	\begin{align}
		T\bar{T}(x) = \lim_{y \to x} \left[T^{\mu\nu}(x)T_{\mu\nu}(y) - T^\mu_\mu(x)T^\nu_\nu(y)\right]\,,
	\end{align}
	leads to a one-parameter family of UV-complete theories with exactly calculable spectra, thermodynamics, and entanglement structures. Remarkably, despite its irrelevant nature, the deformation preserves the modular invariance of the undeformed theory \cite{Cardy:2018sdv,Dubovsky:2018bmo,Datta:2018thy,Aharony:2018bad}, and maintains the integrable structure, including an infinite set of conserved charges. This integrability enables exact computations of a wide range of physical observables, such as the deformed energy spectrum, partition function, scattering $S$-matrix, as well as entanglement and correlation measures \cite{Zamolodchikov:2004ce, Cavaglia:2016oda, Smirnov:2016lqw, He:2019vzf, He:2020cxp,Tian:2023fgf, He:2023obo, He:2023xnb, He:2024pbp}.
	
	From a holographic standpoint, several dual descriptions of $\TTbar$-deformed CFTs have been proposed. The most well-studied realization is the so-called finite-cutoff prescription, where the dual CFT lives on a finite radial slice of AdS$_3$ \cite{McGough:2016lol,Kraus:2018xrn,Taylor:2018xcy,Kraus:2022mnu}, effectively introducing a Dirichlet wall in the bulk\footnote{See, for example,   \cite{Asrat:2017tzd,Shyam:2017znq,Cottrell:2018skz,Hartman:2018tkw,Shyam:2018sro,Jafari:2019qns,Caputa:2019pam,Lewkowycz:2019xse,Giveon:2017myj,Chang:2024voo,Pant:2024eno,Basu:2024enr,FarajiAstaneh:2024fig} for further investigations in this direction.}. An alternative and often more flexible approach involves modifying the boundary conditions of the bulk metric; in the mixed (Dirichlet–Neumann) boundary condition prescription \cite{Guica:2019nzm}, the deformation is encoded through double-trace sources that preserve the conformal boundary while implementing the irrelevant flow. The latter picture is particularly well-suited for accommodating quantum anomalies and allows for a more refined holographic dictionary. Beyond these, other dual holographic avatars have been explored in the literature. These include interpretations of the $\TTbar$ deformation as coupling the CFT to Jackiw-Teitelboim-type topological gravity in two dimensions \cite{Dubovsky:2017cnj,Dubovsky:2018bmo}, realizations via non-critical string theory \cite{Callebaut:2019omt,Tolley:2019nmm}, and the glue-on AdS holography \cite{Apolo:2023vnm,Apolo:2023ckr}. Additionally, connections to random geometry approaches have been proposed, where the $\TTbar$ deformation is interpreted as inducing a stochastic metric on the boundary \cite{Cardy:2018sdv}, with corresponding bulk duals constructed in \cite{Hirano:2020nwq,Hirano:2020ppu}. A comprehensive overview of the historical developments and recent advances in the study of the $\TTbar$ deformation and its holographic interpretation can be found in the extensive reviews \cite{Jiang:2019epa, He:2025ppz}.
	
	While much of the existing literature on $\TTbar$ focuses on parity-symmetric CFTs (with equal left and right-moving central charges, $c^{}_L = c^{}_R$), many physically relevant systems, including the chiral edge modes in quantum Hall systems \cite{Wen:1992vi,Read:2008rn}, heterotic string worldsheet theories \cite{Gross:1985fr}, and the dual CFTs in Kerr/CFT correspondence \cite{Castro:2010fd, Detournay:2012pc}, feature \emph{gravitational anomalies} characterized by $c^{}_L \neq c^{}_R$ \cite{Alvarez-Gaume:1983ihn,Alvarez-Gaume:1984zlq}. Such gravitational anomalies, also appear at the boundaries of topological ordered phases with broken parity \cite{Kitaev:2005dm, Ryu:2012he}, and in thermal transport phenomena governed by anomalous hydrodynamics \cite{Jensen:2012kj}. The natural gravitational dual to such anomalous CFT$_2$s is provided by \emph{topological massive gravity} (TMG) \cite{Deser:1982vy,Li:2008dq}, where a gravitational Chern–Simons term supplements the Einstein–Hilbert action, leading to propagating, parity-violating massive gravitons. This asymmetry provides an excellent setting to explore how irrelevant deformations such as $\TTbar$ interact with chiral anomalies and parity-odd dynamics in the bulk. Studying the $\TTbar$ deformation in anomalous CFT$_2$ via TMG opens up several intriguing questions:
	\begin{itemize}
		\item \textbf{Deformation flow and anomaly structure:} Can the exact solvability and universal deformation of the energy spectrum be preserved when the theory exhibits a gravitational anomaly? In particular, how are the trace flow equations and stress tensor modified under the influence of the anomaly? Does the density of states in the UV also exhibits the Hagedorn behavior\footnote{It seems the Hagedorn behavior is common in $\TTbar$ deformed theories, see \cite{Jiang:2020nnb} for its emergence in $\TTbar$ deformed 1d Bose gas.}?
		
		\item \textbf{Holography with mixed boundary conditions:} In TMG, the presence of the gravitational Chern–Simons term breaks bulk parity and complicates the variational principle. The standard finite-cutoff holography fails in this context, motivating the need for a consistent mixed-boundary prescription \cite{Guica:2019nzm,Tian:2024vln}. It is essential to understand whether this prescription can consistently reproduce the $\TTbar$ flow in theories with $c_L \neq c_R$.
		
		\item \textbf{Quantum information measures:} Entanglement entropy, reflected entropy, and related quantities such as modular flow and entanglement wedge cross-section are sensitive to chiral anomalies. In particular, gravitational anomalies lead to a anomaly-dependence in the holographic probes via spinning worldlines \cite{Castro:2014tta,Jiang:2019qvd,Basu:2022nds,Wen:2022jxr,Wen:2024muv}, or specific geodesic chords settled on the so-called inner RT surface \cite{Wen:2024muv}. It remains nontrivial to determine whether the field-theoretic and holographic computations remain consistent under the influence of $\TTbar$ deformation. 	
	\end{itemize}

	In this work, we develop a holographic framework for the $\TTbar$ deformation of CFT$_2$ with $c^{}_L \neq c^{}_R$, using topological massive gravity as the bulk dual\footnote{Interestingly, heterotic string backgrounds offer an alternative holographic realization of single-trace $\TTbar$ deformations with unequal central charges \cite{Chang:2023kkq,Dei:2024sct}. For massive gravity generalizations of $\TTbar$ deformation in a slightly different context, see \cite{Tsolakidis:2024wut}.}. We construct the deformed BTZ geometry in TMG with mixed boundary conditions and extract the corresponding map between undeformed and deformed parameters. We derive the deformed energy spectrum and demonstrate that the universal flow equation remains valid even when $c^{}_L \neq c^{}_R$, providing strong evidence for solvability. We compute the leading $O(\mu)$ corrections to entanglement and reflected entropy on the twisted cylinder using conformal perturbation theory, and observe perfect agreement with holographic computations involving spinning worldlines in the bulk. We identify a critical bound on the deformation parameter $\mu$ beyond which the holographic EE becomes complex, analogous to a Hagedorn transition, and generalize this to include gravitational anomalies.

	The paper is structured as follows. Section~\ref{review} reviews the salient features of the $\TTbar$ deformation and topological massive gravity. In Section~\ref{sec:perturbation}, we investigate the entanglement and correlation structure through analyzing the leading order corrections to the entanglement entropy and the reflected entropy in a $\TTbar$-deformed anomalous CFT$_2$. Section~\ref{sec:Holography} is devoted to the construction of the deformed BTZ geometry with mixed boundary conditions and the deformed energy spectrum. In Section \ref{sec:HEE}, we compute the holographic entanglement entropy and the entanglement wedge cross-section (EWCS) and compare the field-theoretic and holographic results. We also investigate the balanced partial entanglement entropy (BPE) in deformed anomalous CFTs and confirm the equivalence with the EWCS. Additionally, the implications of the Hagedorn bound on entanglement and correlation are also investigated. Finally, in Section \ref{sec:summary}, we summarize our results and comment on possible future directions.

	\section{Review}\label{review}
	
	\subsection{$\ttbar$ deformation}\label{sec:TTbar-review}
	
	In this subsection, we briefly review the basic features of the $\TTbar$-deformed CFT$_2$. The $\TTbar$ deformation is a universal irrelevant deformation of any $2d$ quantum field theory through the determinant of the stress tensor \cite{Zamolodchikov:2004ce,Cavaglia:2016oda,Smirnov:2016lqw,McGough:2016lol,Guica:2019nzm}:
	\begin{align}\label{OTTbar}
		\CO_{\TTbar}=\gamma_{\alpha\beta}T^{\alpha\eta}T^\beta_{~\eta}-\left(T^\alpha_{~\alpha}\right)^2\,,
	\end{align}
	where $\gamma_{\alpha\beta}$ is the background metric, and $T_{\alpha\beta}$ is the stress tensor of the quantum field theory we study, which is obtained as the response of the action functional to arbitrary variations in the background metric. Such theories obey the following flow equation for the action functional in Euclidean signature\footnote{We follow the conventions in \cite{Guica:2019nzm}.},
	\begin{align}
		\del_\mu \CS^{[\mu]}=-\frac{1}{2}\int\d^2x\,\sqrt{\gamma^{[\mu]}}\,\CO_{\TTbar}^{[\mu]}\,,\label{flow-equation}
	\end{align}
	where $\mu$ is the deformation parameter and the superscripts correspond to the deformed theory. Through the variational principle and utilizing the definition of the instantaneous stress tensor as the response of the action under variations of the background metric, it was shown in \cite{Guica:2019nzm} that the flow equation reduces to the following differential equations
	\begin{equation}
		\partial_\mu\gamma^{\left[\mu\right]}_{\alpha\beta}=-2\hat{T}^{\left[\mu\right]}_{\alpha\beta},\quad \partial_\mu\hat{T}_{\alpha\beta}^{\left[\mu\right]}=-\hat{T}^{\left[\mu\right]}_{\alpha\eta}\,\hat{T}^{\left[\mu\right]\eta}{}_{\beta}\,.
	\end{equation}
	The analytical solutions, in terms of the undeformed metric and stress tensor $(\gamma_{\alpha\beta}^{[0]}, \hat{T}_{\alpha\beta}^{[0]})$, terminate at the second order in the deformation parameter $\mu$ \cite{Guica:2019nzm},
	\begin{align}
		\gamma^{\left[\mu\right]}_{\alpha\beta}=&\:\gamma_{\alpha\beta}^{\left[0\right]}-2\mu\,\hat{T}_{\alpha\beta}^{\left[0\right]}+\mu^2\hat{T}_{\alpha\rho}^{\left[0\right]}\,\hat{T}_{\sigma\beta}^{\left[0\right]}\,\gamma^{\left[0\right]\rho\sigma},\label{flowsolutions}\\
		\hat{T}^{\left[\mu\right]}_{\alpha\beta}=&\:\hat{T}_{\alpha\beta}^{\left[0\right]}-\mu\,\hat{T}_{\alpha\rho}^{\left[0\right]}\,\hat{T}^{\left[0\right]}_{\sigma\beta}\,\gamma^{\left[0\right]\rho\sigma}.\label{deformedst}
	\end{align}
	In the above expressions, $\hat{T}^{\left[\mu\right]}_{\alpha\beta}$ is the (deformed) trace reversed stress tensor defined as,
	\begin{equation}\label{reversedstd}
		\hat{T}^{\left[\mu\right]}_{\alpha\beta}\equiv T_{\alpha\beta}^{\left[\mu\right]}-\gamma_{\alpha\beta}^{\left[\mu\right]}T^{\left[\mu\right]\eta}{}_{\eta}
		\quad
		\Leftrightarrow
		\quad T^{\left[\mu\right]}_{\alpha\beta}=\hat{T}^{\left[\mu\right]}_{\alpha\beta}-\gamma^{\left[\mu\right]}_{\alpha\beta}\hat{T}^{\left[\mu\right]\eta}{}_{\eta}.
	\end{equation}
	In particular, if one considers a conformal field theory as the seed theory of the flow, then the initial condition on the trajectory in the theory space is given by:
	\begin{align}
		\CS^{[\mu]}\Big|_{\mu=0}=\CS_\textrm{CFT}\,,
	\end{align}
	with a symmetric and traceless stress tensor.
	
	\subsection{The holographic description of the
	 $\TTbar$-deformed CFT$_2$}\label{sec holo}
	 
	The cutoff prescription \cite{McGough:2016lol} fundamentally relies on the identification of the induced metric on a specific radial slice with the background metric of the deformed field theory, up to a global constant conformal factor. However, as we will demonstrate, this condition cannot be fulfilled in the context of the topological massive gravity (TMG) \cite{Deser:1982vy,Deser:1981wh}, the gravity theory we focus on in this paper. This indicates that the cutoff prescription could not apply to TMG as observed first in \cite{Tian:2024vln}. Consequently, we adopt the mixed boundary condition prescription proposed in \cite{Guica:2019nzm} as our tool in this paper.  
	
	Let us briefly introduce the principles of the mixed boundary condition prescription. We begin with an auxiliary Ba\~{n}ados geometry \cite{Banados:1998gg}, which describes the general AdS$_3$ solutions with a flat boundary. It can be written in the Fefferman-Graham gauge\footnote{We set the AdS radius $\ell = 1$ in this paper.},
	\begin{equation}\label{Banados geometry}
		ds^2=\frac{d\rho^2}{4\rho^2}+\frac{dudv}{\rho}+\mathcal{L}\left(u\right)du^2+\bar{\mathcal{L}}\left(v\right)dv^2+\rho\mathcal{L}\left(u\right)\bar{\mathcal{L}}\left(v\right)dudv,
	\end{equation}
	where $\mathcal{L}(u)$ and $\bar{\mathcal{L}}(v)$ are two arbitrary chiral functions, and $(u,v)$ are the boundary null coordinates of some auxiliary CFT$_2$, defined on the flat background, i.e.,
	$g^{(0)}_{\alpha\beta}= \eta_{\alpha\beta}$. Upon performing holographic renormalization \cite{deHaro:2000vlm,Balasubramanian:1999re}, one finds that the holographic stress tensor is given by (see Appendix \ref{Sec F-G gauge} for more details),
	\begin{equation}\label{hst}
		g^{\left(2\right)}_{\alpha\beta}=8\pi G\,\hat{T}^{\left[0\right]}_{\alpha\beta}=8\pi G\begin{pmatrix}
			&\CL(u) &0\\
			&0 &\bar\CL(v)
		\end{pmatrix}\,.
	\end{equation}
	Substituting the above relation into \eqref{flowsolutions}, the deformed metric is given by,
	\begin{equation}
		\gamma_{\alpha\beta}^{\left[\mu\right]}dx^\alpha dx^\beta=\left(du-2\mu\bar{\mathcal{L}}\left(v\right)dv\right)\left(dv-2\mu\mathcal{L}\left(u\right)du\right).
	\end{equation}
	Note that we have rescaled the deformation parameter for simplicity, i.e., $\mu \rightarrow \mu/8\pi G$, and keep this convention om the gravity side throughout this paper\footnote{Note that, on the field theory side, we do not rescale the deformation parameter.}. The mixed boundary condition prescription \cite{Guica:2019nzm} then dictates a field-dependent coordinate transformation to manifest \eqref{flowsolutions},
	\begin{equation}\label{dynamical ct}
		dUdV=\left(du-2\mu\bar{\mathcal{L}}\left(v\right)dv\right)\left(dv-2\mu\mathcal{L}\left(u\right)du\right),
	\end{equation}
	where $\left(U,V\right)$ are the two null coordinates of the $\TTbar$-deformed system, given by,
	\begin{equation}\label{dynamical transformation}
		U=u-2\mu\int^{v}\bar{\mathcal{L}}\left(v'\right)dv',\quad V=v-2\mu\int^{u}\mathcal{L}\left(u'\right)du',
	\end{equation}
	where the integrals start at some arbitrary points. By performing these field-dependent coordinate transformations, we can obtain the $\TTbar$-deformed stress tensor and the corresponding $\TTbar$-deformed AdS$_3$ geometry. 
	
	In this work, we mainly focus on the CFT$_2$ at finite temperature with a conserved angular momentum, i.e. the CFT$_2$ defined on a “twisted cylinder” \cite{Caputa:2013lfa}. This setup has the rotating BTZ black hole as its gravity dual. The $\TTbar$-deformed black hole geometry is obtained by performing the field-dependent coordinate transformation \eqref{dynamical transformation} on \eqref{Banados geometry} with two constant parameters $\mathcal{L}(u),\bar{\mathcal{L}}(v)=\mathcal{L}_\mu,\bar{\mathcal{L}}_\mu$. Note that the parameters $\mathcal{L}_\mu,\bar{\mathcal{L}}_\mu$ differ from the undeformed ones $\mathcal{L}_0,\bar{\mathcal{L}}_0$. There are two methods to find the relations between the deformed parameters $\mathcal{L}_\mu,\bar{\mathcal{L}}_\mu$ and the undeformed ones $\mathcal{L}_0,\bar{\mathcal{L}}_0$ \cite{Guica:2019nzm}. The first one relies on the fact that the $\TTbar$ deformation smoothly changes the energy spectrum, and does not affect the local degeneracy of states. Hence, the thermal entropy (i.e., the area of the outer horizon) remains unchanged. Additionally, the angular momentum in the bulk also remains unchanged, since it is quantized and cannot depend on the continuous deformation parameter. The second method involves finding a suitable coordinate transformation to the standard BTZ form that preserves the periodicity of the spatial coordinate. Ultimately, the relations read,
	\begin{equation}\label{relationswithout}
		\begin{aligned}
			\mathcal{L}_\mu=&\frac{1+4\mu\left(\bar{\mathcal{L}}_0+\mu\left(\mathcal{L}_0-\bar{\mathcal{L}}_0\right)^2\right)-\left(1-2\mu\left(\mathcal{L}_0-\bar{\mathcal{L}}_0\right)\right)\sqrt{1+4\mu\left(\mathcal{L}_0+\bar{\mathcal{L}}_0+\mu\left(\mathcal{L}_0-\bar{\mathcal{L}}_0\right)^2\right)}}{8\mu^2\mathcal{L}_0},\\
			\bar{\mathcal{L}}_\mu=&\frac{1+4\mu\left(\mathcal{L}_0+\mu\left(\mathcal{L}_0-\bar{\mathcal{L}}_0\right)^2\right)-\left(1+2\mu\left(\mathcal{L}_0-\bar{\mathcal{L}}_0\right)\right)\sqrt{1+4\mu\left(\mathcal{L}_0+\bar{\mathcal{L}}_0+\mu\left(\mathcal{L}_0-\bar{\mathcal{L}}_0\right)^2\right)}}{8\mu^2\bar{\mathcal{L}}_0}.
		\end{aligned}
	\end{equation}
	Substituting these relations as well as \eqref{hst} into the deformed stress tensor \eqref{deformedst}, we can obtain the (rescaled) deformed energy and angular momentum,
	\begin{equation}\label{deformed EJ}
		\begin{aligned}
			E_{\mu}=&8\pi G\: T^{\left[\mu\right]}_{TT}=\frac{1}{2\mu}\left[-1+\sqrt{1+4\mu\left(\mathcal{L}_0+\bar{\mathcal{L}}_0+\left(\mathcal{L}_0-\bar{\mathcal{L}}_0\right)^2\mu\right)}\right],\\ J_\mu=&8\pi G \:T^{\left[\mu\right]}_{T\phi}=\mathcal{L}_0-\bar{\mathcal{L}}_0,
		\end{aligned}
	\end{equation}
	where $\left(T,\phi\right)$ are the temporal and spatial coordinates, respectively, of the deformed system, i.e. $T=\left(U-V\right)/2,\phi=\left(U+V\right)/2$. The undeformed charges $\left(E,J\right)$ can be obtained by taking the undeformed limit $\mu\rightarrow 0$,
	\begin{equation}\label{decharges}
		E=\mathcal{L}_0+\bar{\mathcal{L}}_0,\quad J=\mathcal{L}_0-\bar{\mathcal{L}}_0.
	\end{equation}
	Substituting these formulas into \eqref{deformed EJ}, we can obtain the deformed energy spectrum,
	\begin{equation}\label{deensp}
		E_\mu=\frac{-1+\sqrt{1+4\mu\left(E+J^2\mu\right)}}{2\mu},
	\end{equation}
	which is consistent with the result of the field theoretical calculations \cite{Smirnov:2016lqw,Cavaglia:2016oda}.

	\subsection{Topological massive gravity and holographic entanglement with gravitational anomalies}
	
	\subsubsection{Topological massive gravity}
	
	In this work, the quantum field theory we study is a two-dimensional CFT$_2$ with unequal central charges in the left and right moving sectors \cite{Alvarez-Gaume:1983ihn,Alvarez-Gaume:1984zlq}. Such theories are called CFT$_2$ with gravitational anomalies (or anomalous CFT$_2$ for short), and there are mainly two approaches to describe them. The first approach has a symmetric stress tensor but not conserved \cite{Kraus:2005zm}, while the second approach has a conserved stress tensor but not symmetric \cite{Alvarez-Gaume:1984zlq}. These two approaches can be related by adding an additional local counterterm to the generating functional \cite{Alvarez-Gaume:1984zlq}. In this paper, we prefer the first approach for two reasons. First, the holographic description of the entanglement entropy for the first approach is more intuitive, corresponding to the worldline action \cite{Castro:2014tta} for a spinning particle in the AdS$_3$ bulk. Second, a symmetric stress tensor is necessary for deriving the deformed metric \eqref{flowsolutions} and the deformed stress tensor \eqref{deformedst}, which serve as our starting point for the holographic $\TTbar$ deformation of the CFT with gravitational anomalies.
	
	The action of the topological massive gravity (TMG) includes the Einstein-Hilbert term and the gravitational Chern-Simons (CS) term \cite{Deser:1982vy,Deser:1981wh},
	\begin{equation}\label{TMG}
		I_{\text{TMG}}=\frac{1}{16\pi G}\int d^3x\sqrt{-g}\left[R-2\Lambda +\frac{1}{2\lambda}\varepsilon^{\alpha\beta\gamma}\left(\Gamma^{\rho}{}_{\alpha\sigma}\partial_\beta \Gamma^{\sigma}{}_{\gamma\rho}+\frac{2}{3}\Gamma^{\rho}{}_{\alpha\sigma}\Gamma^{\sigma}{}_{\beta\eta}\Gamma^{\eta}{}_{\gamma\rho}\right)\right],
	\end{equation}
	where $\Lambda = -1$ is the cosmological constant with the AdS$_3$ radius $\ell = 1$, and  $\lambda$ is the coupling constant that characterizes the interaction strength between the CS term and the Einstein-Hilbert term. The Brown-Henneaux symmetry analysis \cite{Brown:1986nw} of TMG on locally AdS$_3$ backgrounds shows that there are two unequal central charges in the asymptotic symmetry algebra \cite{Kraus:2005zm,Hotta:2008yq,Compere:2008cv},
	\begin{equation}\label{twocentreal}
		c^{}_L=\frac{3}{2G}\left(1+\frac{1}{\lambda}\right), \quad c^{}_R=\frac{3}{2G}\left(1-\frac{1}{\lambda}\right)\,,
	\end{equation}
	indicating a putative duality with the anomalous CFT$_2$.
	Since the central charges represent the degrees of freedom of the system, we naturally require that both $c^{}_L$ and $c^{}_R$ are positive, i.e., $\lambda < -1$ or $\lambda > 1$. The equation of motion of TMG is given by \cite{Castro:2014tta},
	\begin{equation}
		R_{\mu\nu}-\frac{1}{2}Rg_{\mu\nu}+\Lambda g_{\mu\nu}=-\frac{1}{\lambda} C_{\mu\nu},
	\end{equation}
	where $C_{\mu\nu}$ is the Cotton tensor. Consequently, TMG also admits locally AdS$_3$ solutions, for which $C_{\mu\nu}$ vanishes.
	
	\subsubsection{Holographic entanglement with gravitational anomalies}
	In this subsection, we briefly review the holographic entanglement with gravitational anomalies, which encompasses both the holographic entanglement entropy and the entanglement wedge cross-section.
	
	\subsection*{Holographic entanglement entropy}
	
	In \cite{Castro:2014tta}, the authors found that the primary operators in the anomalous CFT$_2$ correspond to massive spinning particles propagating along the extremal worldline in the AdS$_3$ bulk. Correspondingly, the holographic entanglement entropy can be obtained by extremizing a so-called worldline action for a spinning particle,
	\begin{equation}\label{worldline action}
		S_{\text{HEE}}=\frac{1}{4G}\int_{\mathcal{C}}d\tau\left(\sqrt{g_{\mu\nu}\dot{X}^\mu\dot{X}^\nu}+\frac{1}{\lambda}\tilde{\mathbf{n}}\cdot\nabla \mathbf{n}\right),
	\end{equation}
	where $\tau$ is the affine parameter and $\dot{X}^\mu$ is the unit tangent vector of the worldline $\mathcal{C}$, which is homologous to the boundary subsystem $A$ under consideration. The auxiliary vectors $(\tilde{\mathbf{n}}, \mathbf{n})$ together with $\dot{X}$ set up a normal frame at each point on the worldline and satisfy the following constraints
	\begin{equation}
		\tilde{\mathbf{n}}^2=1,\quad \mathbf{n}^2=-1,\quad \tilde{\mathbf{n}}\cdot \mathbf{n}=\tilde{\mathbf{n}}\cdot \dot{X}=\mathbf{n}\cdot \dot{X}=0,
	\end{equation}
	
	For locally AdS$_3$ geometries, the saddle point of this worldline action (i.e., the worldline $\mathcal{C}$) is precisely the RT surface \cite{Castro:2014tta}. Hence, the first term in \eqref{worldline action}, representing the contribution from the Einstein-Hilbert term, captures the length of the RT surface. We refer to this as the normal contribution $S^n_A$. On the other hand, to evaluate the second term, representing the correction from the CS term, it is customary to introduce a reference parallel transported normal frame $\{\mathbf{q},\tilde{\mathbf{q}}\}$ (see \cite{Castro:2014tta} for the explicit construction).
	With the reference parallel transported normal frame $\{\mathbf{q},\tilde{\mathbf{q}}\}$, the integrand in the second term \cite{Castro:2014tta} can be expressed as a total derivative,
	\begin{equation}
		\tilde{\mathbf{n}}\cdot\nabla \mathbf{n}=\partial_\tau\log\left(\left(\mathbf{q}-\tilde{\mathbf{q}}\right)\cdot \mathbf{n}\right),
	\end{equation}
	and the anomalous contribution $S_A^a$ depends only on the normal frame at the two endpoints $(i,f)$ of the worldline,
	\begin{equation}\label{anomalous term}
		S^a_A=\frac{1}{4\lambda G}\log\left(\frac{\left(\mathbf{q}_f-\tilde{\mathbf{q}}_f\right)\cdot \mathbf{n}_f}{\left(\mathbf{q}_i-\tilde{\mathbf{q}}_i\right)\cdot \mathbf{n}_i}\right).
	\end{equation}
	In other words, $S_A^a$ is determined solely by the twist of $\mathbf{n}_f$ relative to $\mathbf{n}_i$, and is independent of normal frame configuration inside $\mathcal{C}$ as well as the choice of the reference normal frame $\{\mathbf{q},\tilde{\mathbf{q}}\}$.
	
	However, the twist of the normal frame is not a dynamical degree of freedom and \eqref{anomalous term} is ``non-covariant'' \cite{Castro:2014tta}, corresponding to the boundary non-covariance associated with gravitational anomaly. More explicitly, one can choose different timelike curves to define the temporary coordinate $\tau$, and use $\partial_\tau$ to define the normal vector $\mathbf{n}$. Then, different choices of the $\tau$ coordinate defines different normal frames along the worldline, resulting in different $S_A^a$ according to \eqref{anomalous term}. When $\mathcal{C}$ is an RT curve anchored on the boundary, it seems natural to choose the normal vectors at the endpoints in the following way \cite{Castro:2014tta},
	\begin{equation}\label{boundary condition}
		\mathbf{n}_i=\mathbf{n}_f\propto \left(\partial_t\right)_{\text{CFT}},
	\end{equation} 
	where $t$ is the time coordinate of the boundary CFT. Under this choice, it was verified in \cite{Castro:2014tta} that the holographic entanglement entropy with gravitational anomalies \eqref{worldline action} matches the result evaluated by the replica trick from the field theory side. For example, for a CFT at finite temperature with a conserved angular momentum (i.e., a CFT with two unequal temperatures in the left and right moving sectors), whose bulk dual is described by the rotating BTZ geometry \eqref{BTZgeometry}, the normal term $S_A^n$ and the anomalous term $S_A^a$ of the holographic entanglement entropy for the interval $A=\left[\left(-\Delta u/2,-\Delta v/2\right),\left(\Delta u/2,\Delta v/2\right)\right]$ are given by,
	\begin{equation}\label{HEEana}
		\begin{aligned}
			S_A^n=&\frac{1}{4G}\text{Length}\left(\mathcal{E}_A\right)=\frac{1}{4G}\log\left[\frac{\beta_u\beta_v}{\pi^2\epsilon^2}\sinh\left(\frac{\pi\Delta u}{\beta_u}\right)\sinh\left(\frac{\pi \Delta v}{\beta_v}\right)\right],\\
			S_A^a=&\frac{1}{4\lambda G}\int_{\mathcal{E}_A}d\tau\: \tilde{\mathbf{n}}\cdot\nabla\mathbf{n}=\frac{1}{4\lambda G}\log\left[\frac{\beta_u \sinh\left(\frac{\pi \Delta u}{\beta_u}\right)}{\beta_v \sinh\left(\frac{\pi \Delta v}{\beta_v}\right)}\right],
		\end{aligned}
	\end{equation}
	where $\mathcal{E}_A$ is the RT surface corresponding to the interval $A$,  and $\epsilon$ is the holographic cut-off. Additionally, $\beta_u$ and $\beta_v$ represent the inverse temperatures of the left and the right moving sectors, respectively, which correspond to the following thermal identification
	\begin{equation}
		\left(u,v\right)\sim \left(u+i\beta_u,v-i\beta_v\right).
	\end{equation}

	\subsection*{Entanglement wedge cross-section}
	Given two bipartite systems $A$ and $B$ with a connected entanglement wedge $\mathcal{W}_{AB}$, the entanglement wedge cross-section (EWCS) \cite{Takayanagi:2017knl,Nguyen:2017yqw} is the saddle geodesic $\Sigma_{AB}$, that anchors on different pieces of the RT surface $\mathcal{E}_{AB}$ of $A \cup B$. In the absence of gravitational anomalies, the length of $\Sigma_{AB}$ was proposed to be the holographic dual of the reflected entropy \cite{Dutta:2019gen}, the balanced partial entanglement entropy (BPE) \cite{Wen:2021qgx,Wen:2022jxr}, or the entanglement of purification \cite{Terhal:2002riz}. Similarly, with gravitational anomalies, it is natural to expect corrections from the CS term analogous to those in the entanglement entropy. Hence the EWCS can again be separated into two terms that represent the contributions from the Einstein-Hilbert term and the CS term, respectively:
	\begin{equation}
		\begin{aligned}
			E_W^n\left(A,B\right)=&\frac{1}{4G}\text{Length}\left(\Sigma_{AB}\right),\\
			E_W^a\left(A,B\right)=&\frac{1}{4\lambda G}\int_{\Sigma_{AB}}d\tau\: \tilde{\mathbf{n}}\cdot \nabla\mathbf{n}=\frac{1}{4\lambda G}\log\left(\frac{\left(\mathbf{q}_f-\tilde{\mathbf{q}}_f\right)\cdot \mathbf{n}_f}{\left(\mathbf{q}_i-\tilde{\mathbf{q}}_i\right)\cdot \mathbf{n}_i}\right),\\
			E_W\left(A,B\right)=&E_W^n\left(A,B\right)+E_W^a\left(A,B\right)
		\end{aligned}
	\end{equation} 
	where $f$ and $i$ represent the two endpoints of $\Sigma_{AB}$. Unfortunately, the endpoints of $\Sigma_{AB}$ are located in the bulk, hence the choice \eqref{boundary condition} does not apply. Naively choosing the time direction as the boundary condition for the timelike normal vector $\mathbf{n}$ is not well motivated, and cannot reproduce the reflected entropy. In \cite{Wen:2022jxr}, the authors proposed a new boundary condition for the spacelike normal vector $\tilde{\mathbf{n}}$, namely $\tilde{\mathbf{n}}$ should be identified with the unit tangent vector to the RT surfaces of $A \cup B$ at the endpoints of $\Sigma_{AB}$\footnote{Furthermore, the authors of \cite{Wen:2024muv} gave an equivalent description for the anomalous contribution using geodesic chords on the so-called inner RT surface, which is the pre-image of the inner horizon in the corresponding Rindler BTZ black string. This means that, like the black hole entropy, the CS correction to the EWCS using the twist of the normal frames also has an origin from the inner horizon. Nevertheless, we will not use the inner RT surface description in this paper.}. Interestingly, this proposal exactly reproduced the field theoretic computations for the reflected entropy and BPE for various mixed state configurations \cite{Basu:2022nds,Wen:2022jxr}. 
	
\begin{figure}
	\centering
	\includegraphics[width=0.7\linewidth]{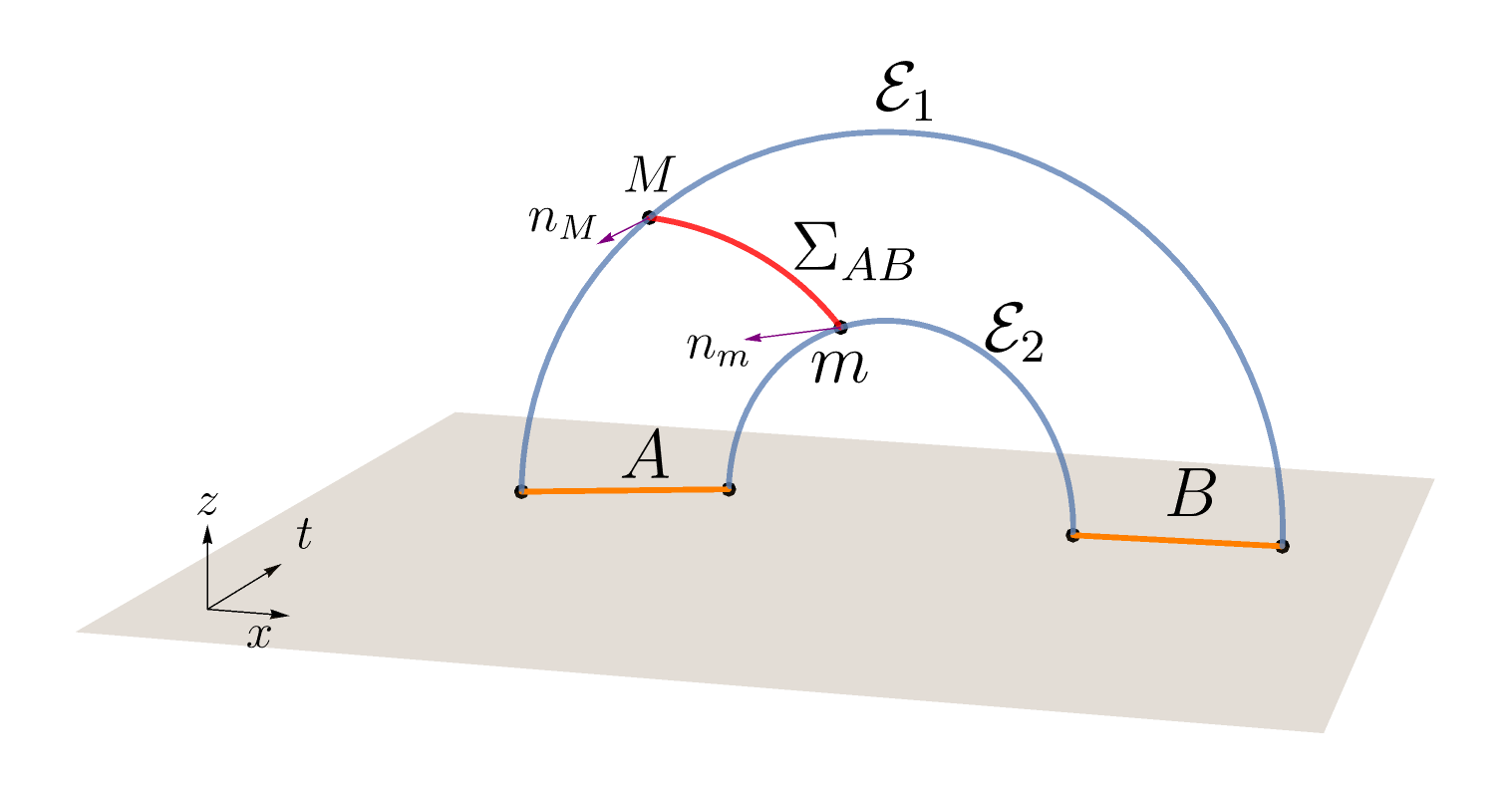}
	\includegraphics[width=0.7\linewidth]{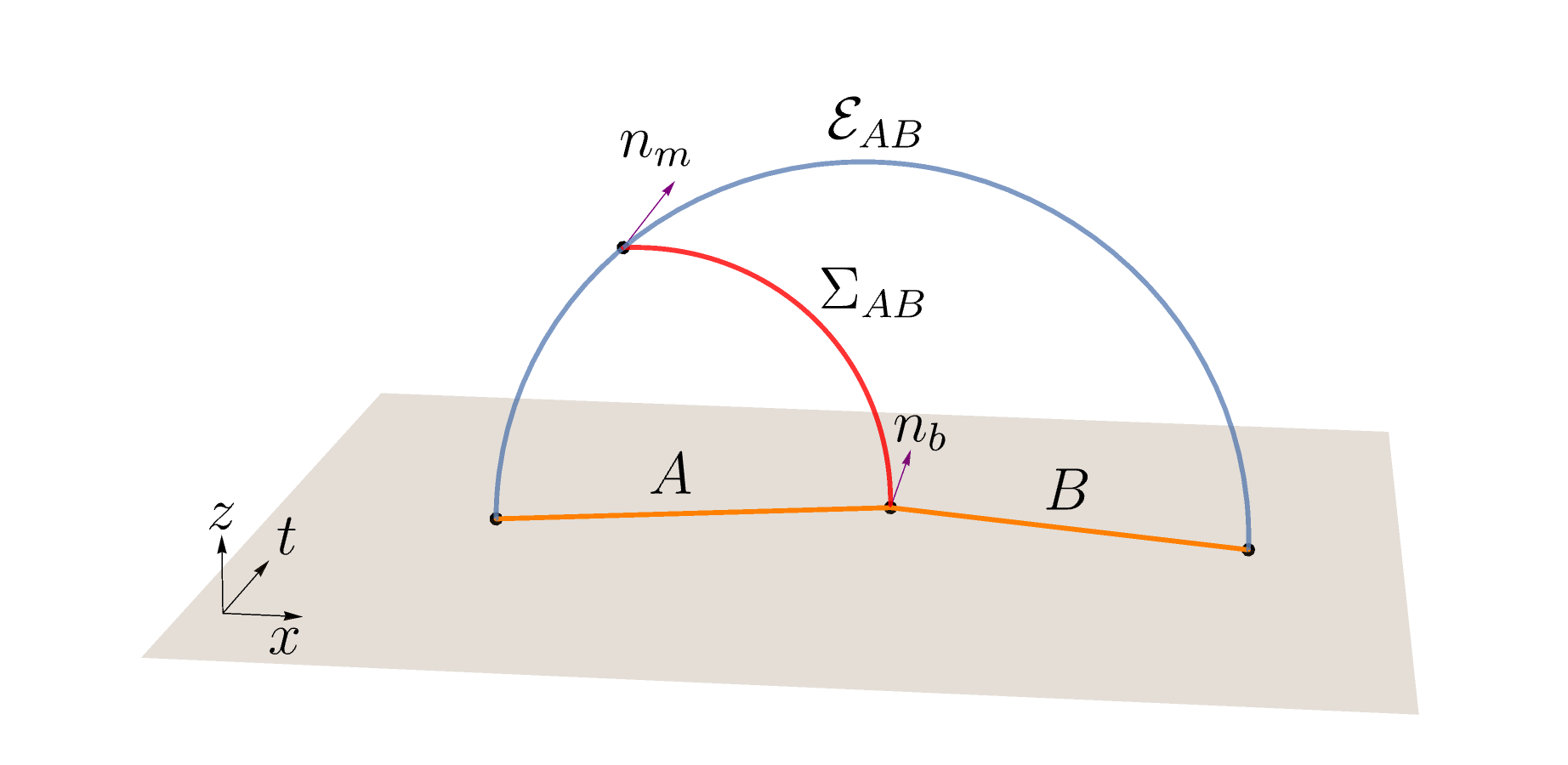}
	\caption{The figures are modified from \cite{Wen:2022jxr}. Illustrations of EWCS in the non-adjacent case and adjacent case, respectively. The red lines represent the saddle geodesic $\Sigma_{AB}$. Top panel: the vectors $n_{M,m}$
	are normal to both $\Sigma_{AB}$ and $\mathcal{E}_{1,2}$. Bottom panel: the vector $n_m$ 
	is normal to both $\Sigma_{AB}$ and $\mathcal{E}_{AB}$, while $n_b\propto \partial_t$.}
	\label{fig:non-adj}
\end{figure}

	More explicitly, consider two covariant non-adjacent intervals $A$ and $B$ on the twisted cylinder as an example,
	\begin{equation}
		A:\left(u_1,v_1\right)\rightarrow \left(u_2,v_2\right),\quad B:\left(u_3,v_3\right)\rightarrow \left(u_4,v_4\right)\,.
	\end{equation}
	The two terms of EWCS are given by \cite{Wen:2022jxr,Basu:2022nds,Wen:2024muv}\footnote{Although the calculation in \cite{Wen:2022jxr} focuses only on the Poincaré AdS$_3$ case, this prescription can be straightforwardly extended to the rotating BTZ geometry and reproduces a half reflected entropy.},
	\begin{equation}\label{nonEWCS}
		\begin{aligned}
			E_W^n\left(A,B\right)=&\frac{1}{8G}\log\left[\frac{\left(\sqrt{\eta}+1\right)\left(\sqrt{\bar{\eta}}+1\right)}{\left(\sqrt{\eta}-1\right)\left(\sqrt{\bar{\eta}}-1\right)}\right],\\
			E_W^a\left(A,B\right)=&\frac{1}{8\lambda G}\log\left[\frac{\left(\sqrt{\eta}+1\right)\left(\sqrt{\bar{\eta}}-1\right)}{\left(\sqrt{\eta}-1\right)\left(\sqrt{\bar{\eta}}+1\right)}\right],
		\end{aligned}
	\end{equation}
	where $\eta$ and $\bar{\eta}$ are two conformal invariants,
	\begin{equation}\label{cratio}
		\eta=\frac{\sinh\left(\frac{\pi u_{12}}{\beta_u}\right)\sinh\left(\frac{\pi u_{34}}{\beta_u}\right)}{\sinh\left(\frac{\pi u_{13}}{\beta_u}\right)\sinh\left(\frac{\pi u_{24}}{\beta_u}\right)},\quad \bar{\eta}=\frac{\sinh\left(\frac{\pi v_{12}}{\beta_v}\right)\sinh\left(\frac{\pi v_{34}}{\beta_v}\right)}{\sinh\left(\frac{\pi v_{13}}{\beta_v}\right)\sinh\left(\frac{\pi v_{24}}{\beta_v}\right)},
	\end{equation}
	with $u_{ij}=u_i-u_j$, and similarly for $v_{ij}$. See \cref{fig:non-adj} for an illustration. In a similar fashion, for two covariant adjacent intervals $A$ and $B$,
	\begin{equation}
		A:\left(u_1,v_1\right)\rightarrow \left(u_2,v_2\right),\quad B:\left(u_2,v_2\right)\rightarrow \left(u_3,v_3\right),
	\end{equation}
	there is an endpoint of $\Sigma_{AB}$ 
	(i.e., the intersection between $A$ and $B$) located at the asymptotic boundary, for which one may still choose the time direction as the boundary condition for the timelike normal vector $\mathbf{n}$ at this endpoint. The two terms of EWCS then read \cite{Wen:2022jxr,Basu:2022nds,Wen:2024muv},
	\begin{small}
		\begin{equation}\label{adjEWCS}
			\begin{aligned}
				E_W^n\left(A,B\right)=&\frac{1}{8G}\log\left[\frac{2\beta_u}{\pi \epsilon}\frac{\sinh\left(\frac{\pi u_{12}}{\beta_u}\right)\sinh\left(\frac{\pi u_{23}}{\beta_u}\right)}{\sinh\left(\frac{\pi u_{13}}{\beta_u}\right)}\right]
				+\frac{1}{8G}\log\left[\frac{2\beta_v}{\pi \epsilon}\frac{\sinh\left(\frac{\pi v_{12}}{\beta_v}\right)\sinh\left(\frac{\pi v_{23}}{\beta_v}\right)}{\sinh\left(\frac{\pi v_{13}}{\beta_v}\right)}\right],\\
				E_W^a\left(A,B\right)=&\frac{1}{8\lambda G}\log\left[\frac{2\beta_u}{\pi \epsilon}\frac{\sinh\left(\frac{\pi u_{12}}{\beta_u}\right)\sinh\left(\frac{\pi u_{23}}{\beta_u}\right)}{\sinh\left(\frac{\pi u_{13}}{\beta_u}\right)}\right]
				-\frac{1}{8\lambda G}\log\left[\frac{2\beta_v}{\pi \epsilon}\frac{\sinh\left(\frac{\pi v_{12}}{\beta_v}\right)\sinh\left(\frac{\pi v_{23}}{\beta_v}\right)}{\sinh\left(\frac{\pi v_{13}}{\beta_v}\right)}\right].
			\end{aligned}
		\end{equation}
	\end{small}
	See \cref{fig:non-adj} for an illustration.
	
	\section{Conformal perturbation theory}\label{sec:perturbation}
	In this section, we investigate the effects of the $\TTbar$ deformation on the entanglement and correlation structure of the anomalous CFT$_2$ at inverse temperature $\beta$ and chemical potential $\Omega$ for the conserved angular momentum $J$, from the field theory side. In the Euclidean signature, the partition function of the seed CFT is given by
	\begin{align}
		Z_\textrm{CFT}=\textrm{Tr}~e^{-\beta (H+i \Omega_E J)}=\textrm{Tr}~e^{-\beta_+L_0-\beta_-\bar L_0}\,,\label{CFT-partition-function}
	\end{align}
	where $\Omega_E=-i\Omega$ is the Wick rotated chemical potential for angular momentum, $(L_0,\bar L_0)$ are the Virasoro zero modes and the temperatures experienced by the left and the right moving modes are defined as follows
	\begin{align}\label{temperatures}
		\beta_\pm=\beta\left(1\pm i\Omega_E\right)\,.
	\end{align}
	This state is characterized by the following expectation values of energy and momentum,
	\begin{align}
		\left<E\right>=E_++E_--\frac{c^{}_L+c^{}_R}{24}~~,~~\left<J\right>=E_+-E_-+\frac{c^{}_L-c^{}_R}{24}
	\end{align}
	featuring non-trivial Casimir energy and momentum for the ground state with $E_\pm=0$. Such a state is conventionally defined on a twisted cylinder\footnote{We focus on a non-compact spatial dimension treating $J$ as linear momentum in our context. We understand this non-compact spatial direction as the limiting case of a very large circle and the resulting boundary topology $\BS^1\times \BR$ corresponds that of a torus with a trivial modular parameter which is referred to as a twisted cylinder in the literature.} with a flat metric
	\begin{equation}
		\d s^2_\textrm{CFT}=-dT^2+d\phi^2=\d\tau^2+\d \phi^2,
	\end{equation}
	where $\tau$ is the Euclidean  time, defined as $\tau=-iT$. The twisted thermal identifications leading to the partition function \eqref{CFT-partition-function} are given by\footnote{In terms of the complex coordinates $w,\bar w:=\phi\pm i\tau$, these identifications correspond to different compactifications of the left and right moving sectors:
		\begin{align}
			(w,\bar w)\sim (w+i\beta_+,\bar{w}+i\beta_-).
	\end{align}}
	\begin{align}
		(\phi,\tau)\sim (\phi-i\beta\Omega_E,\tau+\beta)\,.
	\end{align}
	For a small deformation parameter $\mu$, one may perturbatively expand the action of the deformed theory \eqref{flow-equation} around the undeformed seed CFT$_2$ to obtain
	\begin{align}
		\CS^{[\mu]}=\CS_\textrm{CFT}-\frac{\mu}{\pi^2}\int_\CM\d^2w\,(T\bar{T}-\Theta^2)\,,\label{leading-order-action}
	\end{align}
	where the stress tensor components correspond to the undeformed theory on the twisted cylinder:
	\begin{align}
		T:=2\pi T_{ww}~~,~~\bar{T}:=2\pi T_{\bar{w}\bar{w}}~~,~~\Theta:=2\pi T_{w\bar{w}}\,,
	\end{align}
	and the integral in \eqref{leading-order-action} is understood as 
	\begin{align}
		\int_{\CM}\d^2w:=\int_{0}^{i\Omega_E}\d \phi\int_{0}^{\beta}\d\tau\,.
	\end{align}
	In the following, we will investigate the entanglement and correlation structure of the deformed theory to the leading order in the deformation parameter $\mu$ utilizing a suitable conformal perturbation theory around the seed CFT$_2$, as developed in \cite{Chen:2018eqk, Asrat:2019end,Basu:2023bov,Basu:2023aqz,Basu:2024bal,Basu:2024enr,Jeong:2019xdr,Asrat:2020uib}.
	\subsection{Entanglement entropy}
	
	We begin with the computation of the entanglement entropy of a subsystem $A$ in the $\TTbar$-deformed anomalous CFT$_2$ on the twisted cylinder $\CM$. As is well known, the entanglement entropy in a QFT may be computed in terms of the R\'enyi entropy. Utilizing the replica technique \cite{Calabrese:2004eu,Calabrese:2009qy}, the R\'enyi entropy may be computed as the weighted partition function $\BZ[\CM_n]$ on a $n$-sheeted replica manifold obtained by cyclically gluing copies of the twisted cylinder $\CM$ along the branch cuts present on the copies of the subsystem $A$:
	\begin{align}
		S_n(\rho_A)=\frac{1}{1-n}\log\frac{\BZ[\CM_n]}{\left(\BZ[\CM]\right)^n}\,,
	\end{align}
	where $\BZ[\CM]$ represents the partition function on the original manifold $\CM$. In the $\TTbar$-deformed QFT, the partition function $\BZ[\CM_n]$ is given by the path integral of the deformed action functional on the replica manifold $\CM_n$. In Euclidean signature, the partition function $\BZ[\CM_n]$ reads
	\begin{align}
		\BZ[\CM_n]=\int_{\CM_n}\CD\Phi\,e^{-\CS^{[\mu]}[\Phi;\CM_n]}\,,
	\end{align}
	where $\Phi$ collectively denotes the field content in the theory. To the leading order in $\mu$, we may expand the above expression utilizing \eqref{leading-order-action} to obtain the following correction to the R\'enyi entropy  \cite{Chen:2018eqk,Jeong:2019xdr}
	\begin{align}
		\delta S_n(\rho_A)=-\frac{\mu}{\pi^2(n-1)}\left(\int_{\CM_n}\left<T\bar{T}\right>_{\CM_n}-n\int_{\CM}\left<T\bar{T}\right>_{\CM}\right)\,,\label{delta-Sn}
	\end{align}
	where we have used the fact, for the CFT on a flat background, the $\Theta$ term in \eqref{leading-order-action} vanishes, and the expectation values can be obtained in the replica undeformed CFT$_2$ as follows
	\begin{align}
		\left<T\bar{T}\right>_{\CM_n}=\frac{\displaystyle\int_{\CM_n}\CD\Phi\,e^{-\CS_\textrm{CFT}[\Phi,\CM_n]}T\bar{T}}{\displaystyle\int_{\CM_n}\CD\Phi\,e^{-\CS_\textrm{CFT}[\Phi;\CM_n]}}
	\end{align}
	We consider a single covariant interval $A=[(\phi_1,T_1),(\phi_2,T_2)]$ in the $\TTbar$-deformed anomalous CFT$_2$. Then the expectation value $\left<T\bar{T}\right>_{\CM}$ on the replica manifold may be computed using the twist operator formalism developed in \cite{Jeong:2019xdr}, as follows
	\begin{align}
		\int_{\mathcal{M}_n}\langle T\overline{T}\rangle_{\mathcal{M}_n} 
		&=\frac{1}{n}\int_\mathcal{M}\frac{\langle T^{(n)}(w)\overline{T}^{(n)}(\bar{w})\sigma_n(w_1,\bar{w}_1)\bar{\sigma}_n(w_2,\bar{w}_2)\rangle_{\mathcal{M}}}{\langle\sigma_n(w_1,\bar{w}_1)\bar{\sigma}_n(w_2,\bar{w}_2)\rangle_{\mathcal{M}}}\label{TTb-Mn}\,,
	\end{align}
	where $T^{(n)}$ denotes the total stress tensor for the $n$ replica fields.
	In the case of a CFT$_2$ with gravitational anomalies, the central charges of the left and the right moving sectors are different, and consequently the conformal dimensions of twist operators for the holomorphic and anti-holomorphic sectors are unequal in the same way.
	\begin{align}
		h_{\sigma_n}=\frac{c^{}_L}{24}\left(n-\frac{1}{n}\right),~~\qquad\bar h_{\sigma_n}=\frac{c^{}_R}{24}\left(n-\frac{1}{n}\right)\,,\label{twist-dimensions}
	\end{align}
	rendering the twist operators  anyonic. The correlation functions appearing in \eqref{TTb-Mn} can be straightforwardly computed using the following conformal coordinate transformations to  map the twisted cylinder to the complex plane $\BC$,
	\begin{align}
		w\to z=e^{\frac{2\pi w}{\beta_+}},~~\qquad\bar w\to \bar z=e^{\frac{2\pi \bar w}{\beta_-}}\,.\label{Conformal-map}
	\end{align}
	The stress tensor components on the replica manifold transform as
	\begin{align}
		T^{(n)}(w)=\left(\frac{2\pi}{\beta_+}\right)^2T^{(n)}(z)-\frac{\pi^2n c^{}_L}{6\beta_+^2},~~\qquad\bar T^{(n)}(\bar w)=\left(\frac{2\pi}{\beta_-}\right)^2\bar T^{(n)}(z)-\frac{\pi^2n c^{}_R}{6\beta_-^2}
	\end{align}
	Utilizing $\left<T^{(n)}(z)\right>_\BC=\left<\bar T^{(n)}(z)\right>_\BC=0$, and subsequently applying the conformal Ward identities \cite{Jeong:2019ylz}, we may obtain the expectation value $\left<T\bar{T}\right>_{\CM_n}$ on the replica manifold $\mathcal{M}_n$ as follows
	\begin{align}
		n\left<T\bar{T}\right>_{\CM_n}=&\frac{1}{\langle\sigma_n(z_1, \bar{z}_1)\bar{\sigma}_n(z_2, \bar{z}_2)\rangle_{\BC}}\left[-\frac{\pi^2nc^{}_L}{6{\beta_+}^2}+\left(\frac{2\pi}{\beta_+}z\right)^2\sum_{j=1}^{2}\left(\frac{h_{\sigma_n}}{(z-z_j)^2}+\frac{1}{z-z_j}\partial_{z_j}\right)\right]\notag\\
		&\times\left[-\frac{\pi^2nc^{}_R}{6{\beta_-}^2}+\left(\frac{2\pi}{\beta_-}\bar{z}\right)^2\sum_{j=1}^{2}\left(\frac{\bar{h}_{\sigma_n}}{(\bar{z}-\bar{z}_j)^2}+\frac{1}{\bar{z}-\bar{z}_j}\partial_{\bar{z}_j}\right)\right]
		\langle\sigma_n(z_1, \bar{z}_1)\bar{\sigma}_n(z_2, \bar{z}_2)\rangle_{\BC}.
		\label{TTb-expectation}
	\end{align}
	On the other hand, the $\TTbar$ expectation value on the original twisted cylinder is given by
	\begin{align}
		\langle T\bar{T}\rangle_{\CM}=\frac{\pi^4c^{}_Lc^{}_R}{36\beta_+^2\beta_-^2}\,.
	\end{align}
	The two-point function of the twist operators in the complex plane $\BC$ is given by,
	\begin{equation}
		\left\langle \sigma_n\left(z_1,\bar{z}_1\right) \sigma_n\left(z_2,\bar{z}_2\right)\right \rangle=\frac{c_n}{\left(z_1-z_2\right)^{2h_{\sigma_n}}\left(\bar{z}_1-\bar{z}_2\right)^{2\bar{h}_{\sigma_n}}}.
	\end{equation}
	Now substituting the above expression into \eqref{TTb-expectation}, and utilizing the conformal dimensions \eqref{twist-dimensions} of the twist operators, we can obtain the leading correction to the entanglement entropy of the interval $A$ in the deformed anomalous CFT$_2$ from \eqref{delta-Sn} as
	\begin{align}
		\delta S(A) 
		=&\frac{\mu\pi^2c^{}_Lc^{}_R}{18(\beta_+ \beta_-)^2}\int_{0}^{i\Omega_E}\d \phi\int_{0}^{\beta}\d\tau \Bigg[\frac{ e^{\frac{4 \pi  (\phi+i \tau )}{\beta_+}} \left(z_1-z_2\right)^2}{ \left(e^{\frac{2 \pi  (\phi+i \tau )}{\beta_+}}-z_1\right)^2 \left(e^{\frac{2 \pi  (\phi+i \tau )}{\beta_+}}-z_2\right)^2}\notag\\&\qquad\qquad\qquad\qquad\qquad\qquad\qquad\qquad+\frac{ e^{\frac{4 \pi  (x-i \tau )}{\beta_-}} \left(\bar{z}_1-\bar{z}_2\right)^2}{ \left(e^{\frac{2 \pi  (\phi-i \tau )}{\beta_-}}-\bar{z}_1\right)^2 \left(e^{\frac{2 \pi  (\phi-i \tau )}{\beta_-}}-\bar{z}_2\right)^2}\Bigg]\,,\label{delta-S-integral}
	\end{align} 
	where we have taken the replica limit $n\to 1$.
	The integrals in \eqref{delta-S-integral}, \eqref{refcor} and \eqref{adjrefcor} have been evaluated in \cite{Basu:2024enr} by utilizing a coordinate rotation to nullify the periodic identification in the spatial direction. Substituting the results in \cite{Basu:2024enr}, we can obtain the leading order correction to the entanglement entropy of the single interval as follows
	\begin{align}
		\delta S(A)=\mu\frac{\pi^2c^{}_Lc^{}_R}{18(\beta_+ \beta_-)^{3/2}}\frac{(\phi_{21}-\Omega\,T_{21})}{\sqrt{1-\Omega^2}}\left[\coth \left(\frac{\pi  (\phi_{21}+ T_{21}) }{\beta_+}\right)+\coth \left(\frac{\pi  (\phi_{21}-T_{21}) }{\beta_- }\right)\right]\label{EE-corrections}
	\end{align}
	where we have used the Wick rotation $\tau=-iT\,,\Omega_E=-i\Omega$. As a simple consistency check, we can reproduce the leading correction to the entanglement entropy in the absence of the gravitational anomaly by simply setting $c^{}_L = c^{}_R$. Furthermore, note that the leading correction vanishes identically in the chiral limits $\lambda = \pm 1$, where one of the central charges vanishes.

	\subsection{Reflected entropy}
	
	Next, we turn our attention to another computable correlation measure suitable for characterizing entanglement and correlation in the mixed states, namely, the reflected entropy \cite{Dutta:2019gen}. For a bipartite state $\rho_{AB}\in \CH_A\otimes\CH_B$, the reflected entropy is defined as the von Neumann entropy of a specific reduced density matrix constructed out of the canonically purified state. More explicitly, we introduce the CPT conjugate copies of subsystems $A$ and $B$, which are denoted as $A^\star$ and $B^\star$. The pure state $\ket{\sqrt{\rho_{AB}}}$ lives in the larger Hilbert space $\CH_A\otimes\CH_B\otimes \CH_{A^\star}\otimes \CH_{B^\star}$, and satisfies the purification condition \cite{Dutta:2019gen}
	\begin{align}
		\textrm{Tr}_{\CH_{A^\star}\otimes\CH_{B^\star}}\ket{\sqrt{\rho_{AB}}}\bra{\sqrt{\rho_{AB}}}=\rho_{AB}\,.
	\end{align} 
	Subsequently, the reflected entropy between $A$ and $B$ is defined as the von Neumann entropy of the reduced density matrix $\rho_{AA^\star}$, which is obtained by tracing out the degrees of freedom of subsystems $B$ and $B^\star$ from $\ket{\sqrt{\rho_{AB}}}$:
	\begin{align}
		S_R(A:B)=S_\textrm{vN}(\rho_{AA^\star})_{\sqrt{\rho_{AB}}}\,.
	\end{align}
	In the following, we will evaluate the reflected entropy in the $\TTbar$-deformed anomalous CFT$_2$ by utilizing the replica technique developed in \cite{Basu:2024bal,Asrat:2020uib}, up to the leading order in conformal perturbation theory. Recall that in the replica technique, one first constructs a $m$-fold replication of the original manifold to define the state $\ket{\psi_m}:=\ket{\rho_{AB}^{m/2}}$ as the canonical purification of the state $\rho_{AB}^m$. Subsequently, to construct the R\'enyi reflected entropy, one performs an additional replication in the R\'enyi index $n$, resulting to a $m\times n$ sheeted replica manifold $\CM_{n,m}$ with branch cuts present along individual copies of the subsystems $A$ and $B$. The R\'enyi reflected entropy $S_n(AA^\star)_{\psi_m}$ for the reduced density matrix $\rho^{(m)}_{AA^\star}=\textrm{Tr}_{\CH_B\otimes\CH_{B^\star}}\ket{\psi_m}\bra{\psi_m}$ is then obtained through a properly weighted partition function on this replica manifold as follows
	\begin{align}
		S_n(AA^\star)_{\psi_m}=\frac{1}{1-n}\log\frac{\mathbb{Z}[\CM_{n,m}]}{\left(\mathbb{Z}[\CM_{1,m}]\right)^n}\,.
	\end{align}
	These partition functions may, in turn, be computed through the correlation function of twist operators inserted at the endpoints of the subsystems $A$ and $B$. Such twist operators, denoted as $\sigma_{g^{}_A}$ or $\sigma_{g^{}_B}$, cyclically sew various copies of the subsystems on the replica manifold, according to certain elements $g^{}_A\,,\,g^{}_B$ of the replica symmetry group $\BS_{nm}$. 
	
	Finally, the reflected entropy between $A$ and $B$ may be recovered by taking the replica limit, where one analytically continues the replica parameters and subsequently takes the limit $n,m\to 1$:
	\begin{align}
		S_R(A:B)=\lim_{n,m\to 1}S_n(AA^\star)_{\psi_m}\,.
	\end{align}
	
	A perturbative technique for computing the leading order correction to the reflected entropy due to the $\TTbar$ deformation was developed in \cite{Basu:2024bal,Asrat:2020uib}. In this framework, the first order correction to the R\'enyi reflected entropy can be computed as follows
	\begin{align}
		\delta S_n (A A^\star)_{\psi_m}=-\frac{\mu}{\pi^2(n-1)} \left(\int_{\mathcal{M}_{n,m}}\langle T \overline{T} \rangle_{\mathcal{M}_{n,m}} - n \int_{\mathcal{M}_{1,m}} \langle T \overline{T} \rangle_{\mathcal{M}_{1,m}} \right)\label{SR-correction}\,,
	\end{align}
	which reduces to the correction to the reflected entropy in the replica limit. The expectation value of the $\TTbar$ operator on the replica manifold can be computed utilizing the twist operator formalism \cite{Asrat:2020uib}
	\begin{align}
		\int_{\mathcal{M}_{n,m}}\langle T\overline{T}\rangle_{\mathcal{M}_{n,m}}
		=&\frac{1}{nm}\int_\mathcal{M}\frac{\langle T^{(nm)}(w)\overline{T}^{(nm)}(\bar{w})\prod_{k}\sigma_k(w_k,\bar{w}_k)\rangle_{\mathcal{M}}}{\langle\prod_{k}\sigma_k(w_k,\bar{w}_k)\rangle_{\mathcal{M}}}\notag\\
		=&\frac{1}{nm}\frac{1}{\langle\prod_{k}\sigma_k(z_k,\bar{z}_k)\rangle_{\BC}}\left[-\frac{\pi^2n m c^{}_L}{6{\beta_+}^2}+\left(\frac{2\pi}{\beta_+}z\right)^2\sum_{k}\left(\frac{h_{\sigma_k}}{(z-z_k)^2}+\frac{1}{z-z_k}\partial_{z_k}\right)\right]\notag\\
		&\times\left[-\frac{\pi^2n mc^{}_R}{6{\beta_-}^2}+\left(\frac{2\pi}{\beta_-}\bar{z}\right)^2\sum_{k}\left(\frac{\bar{h}_{\sigma_k}}{(\bar{z}-\bar{z}_k)^2}+\frac{1}{\bar{z}-\bar{z}_k}\partial_{\bar{z}_k}\right)\right]
		\langle\prod_{k}\sigma_k(z_k,\bar{z}_k)\rangle_{\BC}\label{TTb-Mnm}
	\end{align}
	where in the last equality, we have used the conformal map \eqref{Conformal-map} to the complex plane $\CR$, and the conformal Ward identities are exploited\footnote{We have used the shorthand notation $\langle\prod_{k}\sigma_k(w_k,\bar{w}_k)\rangle_\CM$ to denote the correlation function of the twist operator relevant to the given configuration of the subsystems $A$ and $B$. Note that, in this notation, $\sigma_k$ collectively denotes the twist operators inserted at the endpoint $w_k$ or $z_k$ and should be understood as either of $\sigma_{g^{}_A}\,,\,\sigma_{g^{}_B}$ or $\sigma_{g^{-1}_Ag^{}_B}$ depending on the given configuration.}. The expectation value $\langle T \overline{T} \rangle_{\mathcal{M}_{1,m}}$ may be obtained by taking the $n\to 1$ limit of the above expression, with the twist operators $\sigma_{g^{}_A}\,,\sigma_{g^{}_B}$ replaced with $\sigma_{g^{}_m}$ \cite{Asrat:2020uib}. 
	
	\subsubsection{Non-adjacent case}
	
	In particular, we first consider two non-adjacent covariant intervals $A=[(\phi_1,T_1),(\phi_2,T_2)]$ and $B=[(\phi_3,T_3),(\phi_4,T_4)]$ in the anomalous CFT$_2$ with $\TTbar$-deformation. 
	In \eqref{TTb-Mnm}, the relevant correlation function of the twist operators inserted at the endpoints of the subsystems, is given, in the large-central charge limit as follows \cite{Dutta:2019gen,Fitzpatrick:2014vua} 
	\begin{align}
		&\log\langle\sigma_{g^{}_A}(z_1)\sigma_{g^{-1}_A}(z_2)\sigma_{g^{}_B}(z_3)\sigma_{g^{-1}_B}(z_4)\rangle\notag\\\approx& -2h_{g^{}_A}\log(1-\eta)-2\bar h_{g^{}_A}\log(1-\bar\eta)+h_{g^{-1}_Ag^{}_B}\log\left(\frac{1+\sqrt{\eta}}{1-\sqrt{\eta}}\right)+\bar h_{g^{-1}_Ag^{}_B}\log\left(\frac{1+\sqrt{\bar\eta}}{1-\sqrt{\bar\eta}}\right)\label{4-point}
	\end{align}
	where
	\begin{align}
		\eta=\frac{(z_1-z_2)(z_3-z_4)}{(z_1-z_3)(z_2-z_4)}~~,~~\bar\eta=\frac{(\bar z_1-\bar z_2)(\bar z_3-\bar z_4)}{(\bar z_1-\bar z_3)(\bar z_2-\bar z_4)}\,,
	\end{align} 
	are the conformal cross-ratios. Note that the dominant contribution to the above four-point function in the partial wave expansion comes from the exchange of the composite twist operator $\sigma_{g^{-1}_A g^{}_B}$ between $\sigma_{g^{-1}_A}$ and $\sigma_{g^{}_B}$, with conformal dimensions $(h_{g^{-1}_Ag^{}_B}\,,\,\bar h_{g^{-1}_Ag^{}_B})$. The conformal dimensions of these twist operators are given by \cite{Dutta:2019gen}
	\begin{align}\label{C-dim}
		&h_{g^{}_A}=h_{g^{}_B}=n h_{g^{}_m}=\frac{nc^{}_L}{24}\left(m-\frac{1}{m}\right)~~,~~h_{g^{-1}_Ag^{}_B}=\frac{2c^{}_L}{12}\left(n-\frac{1}{n}\right)\,,\notag\\
		&\bar h_{g^{}_A}=\bar h_{g^{}_B}=n \bar h_{g^{}_m}=\frac{nc^{}_R}{24}\left(m-\frac{1}{m}\right)~~,~~\bar h_{g^{-1}_Ag^{}_B}=\frac{2c^{}_R}{12}\left(n-\frac{1}{n}\right)\,.
	\end{align}
	The four-point twist correlator on the $m$-sheeted replica manifold may be readily obtained from that on the $nm$-sheeted surface as
	\begin{align}
		\log\langle\sigma_{g^{}_m}(z_1)\sigma_{g^{-1}_m}(z_2)\sigma_{g^{}_m}(z_3)\sigma_{g^{-1}_m}(z_4)\rangle=\lim_{n \to 1}\log\langle\sigma_{g^{}_A}(z_1)\sigma_{g^{-1}_A}(z_2)\sigma_{g^{}_B}(z_3)\sigma_{g^{-1}_B}(z_4)\rangle\,.
	\end{align}
	Now substituting the four point function \eqref{4-point} and the conformal dimensions of the twist operators \eqref{C-dim} into \eqref{TTb-Mnm} and subsequently employing the replica limit $n,m\to 1$ into \eqref{SR-correction}, we arrive at
	\begin{align}\label{refcor}
		\delta S_R \left(A:B\right)= \frac{\pi^2 c^{}_Lc^{}_R \mu}{9 \beta_+^2\beta_-^2}\int_{0}^{i\Omega_E}\d \phi\int_{0}^{\beta}\d\tau\left( \frac{z^2 \sqrt{(z_1-z_2) (z_1-z_3) (z_2-z_4) (z_3-z_4)}}{(z-z_1) (z-z_2) (z-z_3) (z-z_4)}+ \text{h.c.} \right)  \,.
	\end{align}
	Substituting the result of this integral obtained in \cite{Basu:2024enr} and subsequently performing the Wick rotation $\tau=-i\,T\,,\Omega_E=-i\,\Omega$, we may now obtain the leading order correction to the reflected entropy between the two intervals $A$ and $B$ in the anomalous deformed CFT$_2$ as follows
	\begin{align}
		\delta S_R(A:B)=&\frac{\pi ^2 c^{}_Lc^{}_R \mu }{36 \beta_-^2 \beta_+^2}\sqrt{\eta}\left(\CP_{12}+\CP_{34}-\CP_{14}-\CP_{23}\right)\notag\\
		&+\frac{\pi ^2  c^{}_Lc^{}_R \mu }{36\beta_-^2 \beta_+^2}\sqrt{\bar\eta}\left(\bar\CP_{12}+\bar\CP_{34}-\bar\CP_{14}-\bar\CP_{23}\right)\,,\label{SR-disj-corrections}
	\end{align}
	where $\left(\eta,\bar\eta\right)$ denote the finite temperature cross-ratios:
	\begin{align}
		\eta=\frac{\sinh \left(\frac{\pi  \left(\phi_{21}+T_{21}\right)}{\beta_+}\right) \sinh \left(\frac{\pi  \left(\phi_{43}+T_{43}\right)}{\beta_+}\right)}{\sinh \left(\frac{\pi  \left(\phi_{31}+T_{31}\right)}{\beta_+}\right) \sinh \left(\frac{\pi  \left(\phi_{42}+T_{42}\right)}{\beta_+}\right)},~~\bar\eta=\frac{\sinh \left(\frac{\pi  \left(\phi_{21}-T_{21}\right)}{\beta_-}\right) \sinh \left(\frac{\pi  \left(\phi_{43}-T_{43}\right)}{\beta_-}\right)}{\sinh \left(\frac{\pi  \left(\phi_{31}-T_{31}\right)}{\beta_-}\right) \sinh \left(\frac{\pi  \left(\phi_{42}-T_{42}\right)}{\beta_-}\right)},\label{cross-ratios}
	\end{align}
	and we have defined the functions,
	\begin{equation}
		\begin{aligned}
			\CP_{ij}=&2\beta(\phi_{ij}-\Omega T_{ij})\coth\left(\frac{\pi(\phi_{ij}+T_{ij})}{\beta_+}\right),~~\\
			\bar\CP_{ij}=&2\beta(\phi_{ij}-\Omega T_{ij})\coth\left(\frac{\pi(\phi_{ij}-T_{ij})}{\beta_-}\right)\label{P-functions}.
		\end{aligned}
	\end{equation}
	
	\subsubsection{Adjacent case}
	
	Next we consider the case of two adjacent intervals $A=[(\phi_1,T_1),(\phi_2,T_2)]$ and $B=[(\phi_2,T_2),(\phi_3,T_3)]$ on the twisted cylinder $\CM$. The relevant twist correlation functions in \eqref{TTb-Mnm} on the complex plane are given as
	\begin{align}
		&\langle \sigma_{g^{}_A} (z_1,\bar{z}_1) \sigma_{g^{-1}_A g^{}_B} (z_2,\bar{z}_2) \sigma_{g^{-1}_B} (z_3,\bar{z}_3) \rangle \label{log-sigma-adj}\notag\\
		=&(2m)^{-2h_{g_A^{-1}g^{}_B}-2\bar h_{g_A^{-1}g^{}_B}}(z_1-z_2)^{-2h_{g^{}_A}}(z_2-z_3)^{-2h_{g^{}_A}}(z_1-z_3)^{2h_{g^{}_A}-2h_{g_A^{-1}g^{}_B}}\times\textrm{h.c.}\\
		&\langle \sigma_{g^{}_m} (z_1,\bar{z}_1) \sigma_{g_m^{-1}} (z_3,\bar{z}_3) \rangle=(z_1-z_2)^{-2h_{g^{}_m}}(\bar z_1-\bar z_2)^{-2\bar h_{g^{}_m}}\,,\label{log-sigma-m-adj}
	\end{align}
	Substituting the above expressions and \eqref{C-dim} into \eqref{TTb-Mnm}, and subsequently taking the replica limit $n,m\to 1$ in \eqref{SR-correction}, we can obtain the leading order correction for the reflected entropy,
	\begin{align}\label{adjrefcor}
		\delta S_R(A:B) = \frac{\pi^2 c^{}_Lc^{}_R \mu}{9 \beta_+^2 \beta_-^2}\int_{0}^{i\Omega_E}\d \phi\int_{0}^{\beta}\d\tau \left( \frac{z^2 (z_2 - z_1) (z_2 - z_3)}{(z - z_1) (z - z_2)^2 (z - z_3)} + \textrm{h.c.} \right)\,.
	\end{align}
    Substituting the result of this integral obtained in \cite{Basu:2024enr} and subsequently performing the Wick rotation $\tau=-i\,T\,,\Omega_E=-i\,\Omega$, we may now obtain the leading order correction to the reflected entropy between the two adjacent intervals $A$ and $B$ in the anomalous deformed CFT$_2$ as follows
	\begin{align}\label{SR-adj-correction}
		\delta S_R\left(A:B\right)=&\frac{\pi^2 c^{}_Lc^{}_R \mu}{36\beta_-^2\beta_+^2}\left(\mathbb{P}_{12}+\mathbb{P}_{23}-\mathbb{P}_{13}\right),
	\end{align}
	where $\mathbb{P}_{ij}=\CP_{ij}+\bar\CP_{ij}$. 
	
	\section{The holographic $\TTbar$ deformation for the anomalous CFT}\label{sec:Holography}
	
	\subsection{The mixed boundary condition for the topological massive gravity}
	
	In this section, we investigate the holographic description of the $\TTbar$-deformed anomalous CFT$_2$ through the mixed boundary condition proposal in \cite{Guica:2019nzm}. The central idea of the mixed boundary condition approach is to find the deformed bulk geometry, i.e. the field-dependent coordinate transformation \eqref{dynamical ct}, by substituting the holographic stress tensor \eqref{hst} into the solution of the flow equation \eqref{flowsolutions}. Following the logic in Sec.\ref{sec holo}, we first introduce an auxiliary Ba\~{n}ados geometry (i.e. the rotating BTZ black hole),
	\begin{equation}\label{auxiliary geometry}
		ds^2 =\frac{d\rho^2}{4\rho^2} + \frac{dudv}{\rho} + \mathcal{L}_\mu du^2 + \bar{\mathcal{L}}_\mu dv^2 + \rho \mathcal{L}_\mu \bar{\mathcal{L}}_\mu dudv,
	\end{equation}
	where the parameters $\left(\mathcal{L}_\mu,\bar{\mathcal{L}}_\mu\right)$ are constants, and $\left(u,v\right)$ are the two null coordinates, defined as $u,v=\varphi\pm t$. This line element may be written in the ADM form,
	\begin{equation}\label{BTZgeometry}
		ds^2=-\frac{\left(r^2-r_+^2\right)\left(r^2-r_-^2\right)}{r^2}dt^2+\frac{r^2}{\left(r^2-r_+^2\right)\left(r^2-r_-^2\right)}dr^2+r^2\left(d\varphi-\frac{r_+r_-}{r^2}dt\right)^2,
	\end{equation} 
	by the following radial coordinate transformation,
	\begin{equation}\label{radialct}
		r=\sqrt{\frac{r_+^2+r_-^2}{2}+\frac{1}{\rho}+\frac{\rho}{16}\left(r_+^2-r_-^2\right)^2}.
	\end{equation}
	The relations between the parameters $\left(\mathcal{L}_\mu,\bar{\mathcal{L}}_\mu\right)$ and the radius of the horizons $\left(r_+,r_-\right)$ read,
	\begin{equation}
		r_+=\sqrt{\bar{\mathcal{L}}_\mu}+\sqrt{\mathcal{L}_\mu},\quad r_-=\sqrt{\bar{\mathcal{L}}_\mu}-\sqrt{\mathcal{L}_\mu}.
	\end{equation}
	The thermal identification (i.e. the temperatures) is given by,
	\begin{equation}\label{nodtem}
		\left(u,v\right)\sim\left(u+i\beta_u,v-i\beta_v\right),\quad \beta_u=\frac{2\pi}{r_+-r_-}=\frac{\pi}{\sqrt{\mathcal{L}_\mu}},\quad \beta_v=\frac{2\pi}{r_++r_-}=\frac{\pi}{\sqrt{\bar{\mathcal{L}}}_\mu}.
	\end{equation}
	According to holographic renormalization, the holographic stress tensor of TMG receives corrections from the CS term \cite{Kraus:2005zm}, and is given by,
	\begin{equation}\label{stress0}
		T_{\alpha \beta}^{\left[0\right]}=\frac{1}{8\pi G}\left[g^{\left(2\right)}_{\alpha \beta}-g^{\left(0\right)}_{\alpha\beta}g^{\left(2\right)}_{\gamma\delta}g^{\left(0\right)\gamma\delta}+\frac{1}{2\lambda}g^{\left(0\right)\gamma \delta}\left(\epsilon_{\delta \beta}g^{\left(2\right)}_{\alpha \gamma}+\epsilon_{\delta\alpha}g^{\left(2\right)}_{\beta\gamma}\right)\right],
	\end{equation}
	where $\epsilon_{\alpha\beta}$ 
	represents the Levi-Civita tensor, with $\epsilon_{t\varphi} = -1$. By substituting the coefficients $g^{(0)}$ and $g^{(2)}$ from \eqref{auxiliary geometry} into the above equation, we can obtain the following explicit expression of the holographic stress tensor,
	\begin{equation}\label{ustTMG}
		\hat{T}_{\alpha\beta}^{\left[0\right]}\left(u,v\right)=T_{\alpha\beta}^{\left[0\right]}\left(u,v\right)=\frac{1}{8\pi G}\text{diag}\left(\kappa\mathcal{L}_\mu,\bar{\kappa}\bar{\mathcal{L}}_\mu\right),
	\end{equation}
	where $\kappa,\bar{\kappa}=1\pm1/\lambda$, and we have used the traceless property of the stress tensor. Substituting this expression into \eqref{flowsolutions}, the field-dependent coordinate transformations can be determined by imposing the mixed boundary conditions,
	\begin{equation}\label{deformedboume}
		\gamma^{\left[\mu\right]}_{\alpha\beta}\:dx^\alpha dx^\beta=\left(du-2\mu\bar{\kappa}\bar{\mathcal{L}}_\mu dv\right)\left(dv-2\mu\kappa \mathcal{L}_\mu du\right)=dUdV,
	\end{equation}
	which implies,
	\begin{equation}\label{coordinatetrans}
		U=u-2\mu\bar{\kappa}\bar{\mathcal{L}}_\mu v,\qquad V=v-2\mu\kappa\mathcal{L}_\mu u,
	\end{equation}
	or
	\begin{equation}\label{coordinatetrans1}
		 u=\frac{U+2\mu\bar{\kappa}\bar{\mathcal{L}}_\mu V}{1-4\mu^2\kappa\bar{\kappa}\mathcal{L}_\mu\bar{\mathcal{L}}_\mu},\qquad v=\frac{V+2\mu\kappa\mathcal{L}_\mu U}{1-4\mu^2\kappa\bar{\kappa}\mathcal{L}_\mu\bar{\mathcal{L}}_\mu}.
	\end{equation}
	Plugging the field-dependent coordinate transformations back into \eqref{auxiliary geometry}, we can obtain the deformed bulk metric,
	\begin{equation}\label{deformed geometry}
		\begin{aligned}
			ds^2=\:\frac{d\rho^2}{4\rho^2}&\:+\frac{\left(dV+2\mu\kappa\mathcal{L}_\mu dU\right)\left(dU+2\mu\bar{\kappa}\bar{\mathcal{L}}_\mu dV\right)}{\rho \left(1-4\mu^2\kappa\bar{\kappa}\mathcal{L}_\mu\bar{\mathcal{L}}_\mu\right)^2}\\
			&\:+\frac{\bar{\mathcal{L}}_\mu\left(dV+2\mu\kappa\mathcal{L}_\mu dU\right)^2+\mathcal{L}_\mu\left(dU+2\mu\bar{\kappa}\bar{\mathcal{L}}_\mu dV\right)^2}{\left(1-4\mu^2\kappa\bar{\kappa}\mathcal{L}_\mu \bar{\mathcal{L}}_\mu\right)^2}\\
			&\:+\rho\mathcal{L}_\mu\bar{\mathcal{L}}_\mu \frac{\left(dV+2\mu\kappa\mathcal{L}_\mu dU\right)\left(dU+2\mu\bar{\kappa}\bar{\mathcal{L}}_\mu dV\right)}{ \left(1-4\mu^2\kappa\bar{\kappa}\mathcal{L}_\mu\bar{\mathcal{L}}_\mu\right)^2}.
		\end{aligned}
	\end{equation}
	
	Before concluding this subsection, we confirm that the cutoff prescription is indeed not applicable in the TMG. The rescaled induced metric of \eqref{auxiliary geometry} on the fixed radial surface $\rho = \rho_c$ is given by,
	\begin{equation}
		h_{\alpha\beta}dx^\alpha dx^\beta=dudv+\rho_c\left(\mathcal{L}_\mu du^2+\bar{\mathcal{L}}_\mu dv^2\right)+\rho_c^2\mathcal{L}_\mu\bar{\mathcal{L}}_\mu dudv.
	\end{equation}
	Equating this equation with \eqref{deformedboume}, we find that there is no consistent solution for a constant $\rho_c$ for generic values of $\kappa$ and $\bar{\kappa}$, which indicates that the cutoff prescription fails in TMG.
	
	\subsection{The deformed energy spectrum}
	
	In the $(U,V)$ coordinate system, the deformed trace-reversed stress tensor can be obtained by substituting the undeformed holographic stress tensor \eqref{ustTMG} into \eqref{deformedst} and then making the field-dependent coordinate transformation \eqref{coordinatetrans},
	\begin{equation}
		\hat{T}_{\alpha\beta}^{\left[\mu\right]}\left(U,V\right)=\frac{1}{8\pi G \left(1-4\mu^2\kappa\bar{\kappa}\mathcal{L}_\mu\bar{\mathcal{L}}_\mu\right)}\begin{pmatrix}
			\kappa\mathcal{L}_\mu & 2\mu\kappa\bar{\kappa}\mathcal{L}_\mu\bar{\mathcal{L}}_\mu \\
			2\mu\kappa\bar{\kappa}\mathcal{L}_\mu\bar{\mathcal{L}}_\mu & \bar{\kappa}\bar{\mathcal{L}}_\mu \\
		\end{pmatrix}.
	\end{equation}
	Using the definition \eqref{reversedstd} of the trace-reversed stress tensor $\hat{T}^{\left[\mu\right]}_{\alpha\beta}$, we can obtain the deformed stress tensor of the anomalous CFT,
	\begin{equation}
		T_{\alpha\beta}^{\left[\mu\right]}\left(U,V\right)=\frac{1}{8\pi G\left(1-4\mu^2\kappa\bar{\kappa}\mathcal{L}_\mu\bar{\mathcal{L}}_\mu\right)}\begin{pmatrix}
			\kappa\mathcal{L}_\mu & -2\mu\kappa\bar{\kappa}\mathcal{L}_\mu\bar{\mathcal{L}}_\mu \\
			-	2\mu\kappa\bar{\kappa}\mathcal{L}_\mu\bar{\mathcal{L}}_\mu & \bar{\kappa}\bar{\mathcal{L}}_\mu \\
		\end{pmatrix}.
	\end{equation}
	Therefore, the (rescaled) deformed energy and angular momentum are,
	\begin{equation}\label{dechtmg}
		\begin{aligned}
			E_\mu=&8\pi GT^{\left[\mu\right]}_{TT}=\frac{\kappa\mathcal{L}_\mu+\bar{\kappa}\bar{\mathcal{L}}_\mu+4\mu\kappa\bar{\kappa}\mathcal{L}_\mu\bar{\mathcal{L}}_\mu}{1-4\mu^2\kappa\bar{\kappa}\mathcal{L}_\mu\bar{\mathcal{L}}_\mu},\\
			J_\mu=&8\pi G T^{\left[\mu\right]}_{T\phi}=\frac{\kappa\mathcal{L}_\mu-\bar{\kappa}\bar{\mathcal{L}}_\mu}{1-4\mu^2\kappa\bar{\kappa}\mathcal{L}_\mu\bar{\mathcal{L}}_\mu},
		\end{aligned}
	\end{equation} 
	where $\left(T,\phi\right)$ are the temporal and spatial coordinates of the deformed system. Similarly, for the corresponding state in the undeformed anomalous CFT, which is dual to a Ba\~{n}ado geometry \eqref{auxiliary geometry} with $\mathcal{L}_\mu=\mathcal{L}_0,\bar{\mathcal{L}}_\mu=\bar{\mathcal{L}}_0$ and $u=U,v=V$, the undeformed energy and angular momentum are simply,
	\begin{equation}\label{undctmg}
		E=\kappa\mathcal{L}_0+\bar{\kappa}\bar{\mathcal{L}}_0,\quad J=\kappa \mathcal{L}_0-\bar{\kappa}\bar{\mathcal{L}}_0.
	\end{equation}
	One can easily check that these charges will reduce to those for non-anomalous CFT given in \eqref{decharges} when taking the non-anomalous limit $\lambda \rightarrow \infty$. 
	
	In order to compare the deformed energy and the original one, we need to find the relations between the parameters $\left(\mathcal{L}_\mu,\bar{\mathcal{L}}_\mu\right)$ and $\left(\mathcal{L}_0,\bar{\mathcal{L}}_0\right)$. Our strategy is to use the two methods which are proposed in \cite{Guica:2019nzm} and reviewed in Sec.\ref{sec holo}. For the first method, the angular momentum and the thermal entropy of the deformed geometry \eqref{deformed geometry} should be equal to the original ones, which we calculate one by one. For the TMG, the thermal entropy is corrected by the CS term accordingly \cite{Kraus:2005zm,Sahoo:2006vz,Park:2006gt,Tachikawa:2006sz},
	\begin{equation}\label{thermalcor}
		S=\frac{\ell_{\text{outer horizon}}}{4G}+\frac{\ell_{\text{inner horizon}}}{4\lambda G},
	\end{equation}
	where $\ell_{\text{outer horizon}}$ and $\ell_{\text{inner horizon}}$ represent the lengths of the outer horizon and the inner horizon respectively. According to the radial coordinate transformation \eqref{radialct}, the two horizons are located at,
	\begin{equation}
		\rho=\pm \rho_h,\quad \rho_h=\left(\mathcal{L}_\mu\bar{\mathcal{L}}_\mu\right)^{-\frac{1}{2}}.
	\end{equation}
	The $g_{\phi\phi}$ component of the deformed geometry \eqref{deformed geometry} reads,
	\begin{equation}
		g_{\phi\phi}=\frac{\left(1+\bar{\mathcal{L}}_\mu\left(\rho+2\mu\left(\bar{\kappa}+\kappa\mathcal{L}_\mu\rho\right)\right)\right)\left(1+\mathcal{L}_\mu\left(\rho+2\mu\left(\kappa+\bar{\kappa}\bar{\mathcal{L}}_\mu \rho\right)\right)\right)}{\rho\left(1-4\mu^2\kappa\bar{\kappa}\mathcal{L}_\mu\bar{\mathcal{L}}_\mu\right)^2}.
	\end{equation}
	We compactify the spatial direction, i.e. $\phi\sim\phi+R$, then the lengths of the horizons for the deformed geometry are,
	\begin{equation}
		\begin{aligned}
			\ell^{\mu}_{\text{outer horizon}}=R\sqrt{g_{\phi\phi}}|_{\rho=\rho_h}=&\frac{\sqrt{\bar{\mathcal{L}}_\mu}\left(1+2\mu\kappa\mathcal{L}_\mu\right)+\sqrt{\mathcal{L}_\mu}\left(1+2\mu\bar{\kappa}\bar{\mathcal{L}}_\mu\right)}{\left(1-4\mu^2\kappa\bar{\kappa}\mathcal{L}_\mu\bar{\mathcal{L}}_\mu\right)}R,\\
			\ell^{\mu}_{\text{inner horizon}}=R\sqrt{g_{\phi\phi}}|_{\rho=-\rho_h}=&\frac{\sqrt{\bar{\mathcal{L}}_\mu}\left(1+2\mu\kappa\mathcal{L}_\mu\right)-\sqrt{\mathcal{L}_\mu}\left(1+2\mu\bar{\kappa}\bar{\mathcal{L}}_\mu\right)}{\left(1-4\mu^2\kappa\bar{\kappa}\mathcal{L}_\mu\bar{\mathcal{L}}_\mu\right)}R.
		\end{aligned}
	\end{equation}
	Substituting these expressions into \eqref{thermalcor}, the thermal entropy of the deformed BTZ geometry \eqref{deformed geometry} is given by,
	\begin{equation}\label{modifiedarea}
		\begin{aligned}
			S_{\mu}=&\frac{\ell^{\mu}_{\text{outer horizon}}}{4G}+\frac{\ell^{\mu}_{\text{inner horizon}}}{4\lambda G}\\
			=&\frac{-\sqrt{\bar{\mathcal{L}}_\mu}\left(2+\kappa-\bar{\kappa}\right)\left(1+2\mu\kappa\mathcal{L}_\mu\right)+\sqrt{\mathcal{L}_\mu}\left(-2+\kappa-\bar{\kappa}\right)\left(1+2\mu\bar{\kappa}\bar{\mathcal{L}}_\mu\right)}{8G\left(-1+4\mu^2\kappa\bar{\kappa}\mathcal{L}\bar{\mathcal{L}}_\mu\right)}.
		\end{aligned}
	\end{equation}
	Similarly, the undeformed thermal entropy can be obtained by taking the undeformed limit, i.e. $\mu\rightarrow 0$ and substituting  $\left(\mathcal{L}_\mu,\bar{\mathcal{L}}_\mu\right)$ with $\left(\mathcal{L}_0,\bar{\mathcal{L}}_0\right)$,
	\begin{equation}\label{originalarea}
		S=\frac{\sqrt{\bar{\mathcal{L}}_0}\left(2+\kappa-\bar{\kappa}\right)+\sqrt{\mathcal{L}_0}\left(2-\kappa+\bar{\kappa}\right)}{8G}.
	\end{equation}
	As we mentioned above, the angular momentum and the thermal entropy of the deformed geometry \eqref{deformed geometry} should remain unchanged. According to \eqref{dechtmg}, \eqref{undctmg}, \eqref{modifiedarea}, and \eqref{originalarea}, the relations between $\left(\mathcal{L}_{\mu}, \bar{\mathcal{L}}_\mu\right)$ and $\left(\mathcal{L}_0, \bar{\mathcal{L}}_0\right)$, which exhibit good behavior in the limit $\mu \rightarrow 0$, are given by
	\begin{equation}\label{parelaTMG}
		\begin{aligned}
			\mathcal{L}_\mu=&\frac{1+4\mu\left(\bar{\kappa}\bar{\mathcal{L}}_0+\chi^2\mu\right)-\left(1-2\mu\chi\right)\sqrt{1+4\mu\left(\bar{\chi}+\chi^2\mu\right)}}{8\mu^2\kappa^2\mathcal{L}_0},\\
			\bar{\mathcal{L}}_\mu=&\frac{1+4\mu\left(\kappa\mathcal{L}_0+\chi^2\mu\right)-\left(1+2\mu\chi\right)\sqrt{1+4\mu\left(\bar{\chi}+\chi^2\mu\right)}}{8\mu^2\bar{\kappa}^2\bar{\mathcal{L}}_0},
		\end{aligned}
	\end{equation}
	where
	\begin{equation}
		\chi=\kappa\mathcal{L}_0-\bar{\kappa}\bar{\mathcal{L}}_0,\quad \bar{\chi}=\kappa\mathcal{L}_0+\bar{\kappa}\bar{\mathcal{L}}_0.
	\end{equation}
	One can easily check that these expressions will reduce to \eqref{relationswithout} when taking the non-anomalous limit, i.e. $\lambda\rightarrow \infty$. Furthermore, we show how to use the second method to obtain the same conclusions in Appendix.\ref{appendix relations}. Combining \eqref{dechtmg}, \eqref{undctmg} with \eqref{parelaTMG}, we can obtain the deformed energy spectrum of the anomalous CFT,
	\begin{equation}
		E_{\mu}=\frac{-1+\sqrt{1+4\mu\left(E+\mu^2J^2\right)}}{2\mu},\label{def-spectrum}
	\end{equation} 
	which is the same as the deformed spectrum \eqref{deensp} for the non-anomalous CFT$_2$ cases. 
	
    Now we argue that, on the field theory side the deformed spectrum for anomalous CFT$_2$ also has the same formula \eqref{deensp} as the non-anomalous CFT$_2$. Note that, the formula \eqref{deensp} is derived from the Zamolodchikov's factorization formula of the $\TTbar$ operator \cite{Zamolodchikov:2004ce},
		\begin{align}
			\langle \TTbar \rangle = \langle T \rangle \langle \bar{T} \rangle - \langle \Theta \rangle^2,
		\end{align}
	which is a result of the translation invariance and the conservation of the stress tensor of the theory. For the case of anomalous CFT$_2$ that duals to the TMG, the rotational symmetries are not preserved in the TMG action \eqref{TMG}, but we still have the translation symmetries. Also, in the presence of gravitational anomaly, the conservation of the stress tensor does not hold due to the curvature of the background metric (see \cite{Castro:2014tta})
	\begin{align}
		\nabla^{i}T_{ij}=\frac{c_L-c_R}{96\pi}\epsilon^{kl}\partial_{k}\partial_{m}\Gamma^{m}_{jl}\,.
	\end{align}
	Nevertheless, in this paper we focus on CFT$_2$ in \textit{flat background}, which also remains flat under the $\TTbar$ flow. In conclusion, in the anomalous CFT$_2$ we consider, the translation invariance is preserved, and the stress tensor remains conserved. Hence, Zamolodchikov's original argument for factorization remains valid, and the associated inviscid Burger's-type differential equation governing the $\TTbar$-deformed energy levels continues to apply, and eventually we will get the same formula \eqref{deensp} for the deformed spectrum in the presence of the gravitational anomaly\footnote{We thank Prof. Jiang Yunfeng for pointing this out.}.
		
	\paragraph{Existence of a maximum temperature:} Interestingly, in \cite{Tian:2024vln}, it was shown that the $\TTbar$-deformed partition function of a holographic anomalous CFT$_2$ is still modular invariant. In particular, the authors in \cite{Tian:2024vln} demonstrated that the holographic on-shell actions of the deformed BTZ black hole and global AdS saddles are related by the $S$-modular transformation 
		\begin{align}
			\tau\to-\frac{1}{\tau}~~,~~\mu\to\frac{\mu}{|\tau|^2}\,,\label{S-transf}
		\end{align}
	where $\tau=\tau_1+i\tau_2$ is the modular parameter specifying the torus on which the deformed anomalous CFT$_2$ is defined. The change in the deformation parameter under the modular transformation implies that $\TTbar$-deformed anomalous CFT$_2$s exhibit a maximum temperature. To illustrate this, note that the ground state energy of the deformed anomalous CFT$_2$ is given by
	\begin{align}
		E_0(\mu)=\frac{1}{2\mu}\left(-1+\sqrt{1+4\mu E_0+4\mu^2 J_0^2}\right)\,,\label{def-ground-state}
	\end{align}
	where the Casimir energy and momentum of the undeformed CFT$_2$ are given by (we work in units where $8\pi G=1$)
	\begin{align}
		E_0=-\frac{c^{}_L+c^{}_R}{24}~~,~~J_0=\frac{c^{}_L-c^{}_R}{24}\,.
	\end{align}
	It is evident that the ground state energy exhibits a square root singularity and becomes imaginary unless the deformation parameter satisfies 
	\begin{align}
		\mu\leq \lambda\left(\lambda\pm\sqrt{\lambda^2-1}\right)\,,
	\end{align}
	where we have used \eqref{twocentreal}. Note that the positivity of the central charges \eqref{twocentreal} implies that the TMG coupling $\lambda$ must satisfy either $\lambda>1$ or $\lambda<-1$. The two signs of the upper bound on $\mu$ in the above expression correspond to the two signatures of the coupling $\lambda$.
		
	Now, under the $S$-modular transformation \eqref{S-transf}, we must have
	\begin{align}
		|\tau|^2\geq \frac{\mu}{\lambda\left(\lambda\pm\sqrt{\lambda^2-1}\right)}\,.
	\end{align}
	For an anomalous CFT$_2$ defined on a torus with modular parameter $2\pi\tau=-\beta\Omega_E+i\beta\equiv i\beta_+$ the above inequality leads to a maximum temperature
	\begin{align}
		\frac{\beta^2(1-\Omega^2)}{4\pi^2}\geq \frac{\mu}{\lambda\left(\lambda\pm\sqrt{\lambda^2-1}\right)}\,.\label{T-max}
	\end{align}
	This may be interpreted as the anomalous extension of the Hagedorn temperature described in \cite{Cavaglia:2016oda,Dubovsky:2012wk,McGough:2016lol} for the case without gravitational anomaly. We will explore further holographic evidences of such a maximum temperature in the context of the quantum information theoretic properties of deformed anomalous CFT$_2$s in subsection \ref{sec:Bounds-on-mu}.
	\subsection{Asymptotic density of states and Hagedorn behavior} \label{Hagedorn}
	In this subsection, we analyze the asymptotic density of states in the $\TTbar$-deformed anomalous CFT$_2$ and demonstrate a Hagedorn like behavior in the UV regime, indicating the breakdown of the theory above the Hagedorn temperature \eqref{T-max}.
	Similar observations have been made in \cite{Datta:2018thy} in the absence of gravitational anomaly, and we closely follow their approach here. Consider a rectangular torus with modular parameter $\tau=\theta+i\beta$ where $\theta=-\beta\Omega_E$. The partition function of the $\TTbar$-deformed anomalous CFT$_2$ is given by
	\begin{align}
		\CZ\left(\tau,\bar\tau\big|\mu\right)=\sum_n e^{2\pi i\tau_1 P_n-2\pi \tau_2 E_n(\mu)}
	\end{align}
	For simplicity, we restrict to the vanishing angular potential case with $\Omega_E=0$; hence the partition function simplifies to 
	\begin{align}
		\CZ(\tau,\bar\tau\big|\mu)=\sum_n e^{-2\pi\beta E_n(\mu)}
	\end{align}
	At low temperature $\beta\to\infty$ the above sum is dominated by the ground state\footnote{Note that there is no level crossing in the deformed energy spectrum \eqref{def-spectrum}.} with energy given by \eqref{def-ground-state}. Under the $S$-modular transformation \eqref{S-transf} we find that
	\begin{align}
		\CZ(\mu)\approx\exp\left[-\frac{\pi\beta}{\mu}\left(-1+\sqrt{1+\frac{4\mu}{\beta^2}E_0+\frac{4\mu^2}{\beta^4}J_0^2}\right)\right]
	\end{align}
	Identifying the above result with
	\begin{align}
		\CZ(\mu)=\int_{E_0(\mu)}^{\infty}\rho(E)e^{-2\pi\beta E}\d E
	\end{align}
	we may obtain the asymptotic density of states by performing the inverse Laplace transformation
	\begin{align}
		\rho(E)=\oint \d\beta\exp\left[\beta E-\frac{\pi\beta}{\mu}\left(-1+\sqrt{1+\frac{4\mu}{\beta^2}E_0+\frac{4\mu^2}{\beta^4}J_0^2}\right)\right]
	\end{align}
	Usually such integrals are performed under a saddle-point approximation \cite{Datta:2018thy}. However, in the present case the saddle-point analysis becomes rather cumbersome and results in a $8$-th order polynomial equation in $\beta$. In order to simplify the analysis, we make the simplifying assumption of a UV limit: $\mu E\gg 1$. In this case, the saddle point takes a simpler form
	\begin{align}
		\beta_\star^2=2\left(E_0\pm\sqrt{E_0^2-J_0^2}\right)\mu+\CO\left(\frac{1}{\mu E}\right)=1\mp\frac{\sqrt{\lambda^2-1}}{\lambda}\mu+\CO\left(\frac{1}{\mu E}\right)\,,
	\end{align}
	which results in the Hagedorn behavior
	\begin{align}
		\rho(E)\sim e^{\beta_\star E}=\exp\left[\left(1\mp\frac{\sqrt{\lambda^2-1}}{\lambda}\mu\right)E\right]
	\end{align}
	We may now identify the anomalous extension of the Hagedorn temperature as follows
	\begin{align}
		\beta_H=2\pi\sqrt{1\mp\frac{\sqrt{\lambda^2-1}}{\lambda}\mu}\,,
	\end{align}
	which is identical to \eqref{T-max} for $\Omega=0$, substantiating our claim. As discussed in \cite{Datta:2018thy}, the Hagedorn behavior also implies singularity of the $\TTbar$-deformed partition function near the Hagedorn temperature.

	\section{Holographic entanglement in the $\TTbar$-deformed geometry}\label{sec:HEE}
	
	In this section, we investigate the $\TTbar$-deformed holographic entanglement entropy and entanglement wedge cross-section by employing the worldline action \eqref{worldline action} in the $\TTbar$-deformed geometry \eqref{deformed geometry}\footnote{The derivation of the worldline action does not depend on whether the Dirichlet boundary conditions or the mixed boundary conditions are applied at the asymptotics \cite{Castro:2014tta}; hence it may be safely assumed that the worldline action is still applicable to the $\TTbar$-deformed BTZ geometry.}. The difference between the deformed geometry \eqref{deformed geometry} and the auxiliary Ba\~{n}ados geometry \eqref{auxiliary geometry} lies in a field-dependent coordinate transformations \eqref{coordinatetrans}. Therefore, we can apply the coordinate transformations \eqref{coordinatetrans} to the known results, such as the holographic entanglement entropy in the auxiliary Ba\~{n}ados geometry, to obtain the physical quantities of the $\TTbar$-deformed system holographically. Meanwhile, we need to replace the parameters of the auxiliary system (which is dual to the auxiliary Ba\~{n}ados geometry), such as the temperatures, with those of the $\TTbar$-deformed system. In the following, we apply this prescription to obtain the $\TTbar$-deformed holographic entanglement entropy and EWCS. With these results, we also analyze physical bounds on the deformation parameter.

	\subsection{Holographic entanglement entropy}
	
	Suppose that the thermal identification of the $\TTbar$-deformed system is given by,
	\begin{equation}
		\left(U,V\right)\sim\left(U+i\beta_{U},V-i\beta_{V}\right),
	\end{equation} 
	where $\left(\beta_U,\beta_V\right) $ correspond to the $\TTbar$-deformed versions of $\left(\beta_+,\beta_-\right)$ in \eqref{temperatures}, respectively. At this point, it is important to emphasize that in \cref{sec:perturbation}, we performed perturbative computations about a seed conformal theory and hence $\beta_\pm$ in \eqref{temperatures} correspond to the left and right-moving temperatures of the undeformed anomalous CFT$_2$. According to the coordinate transformations \eqref{coordinatetrans}, the relations between the inverse temperatures $\left(\beta_U,\beta_V\right)$ of the $\TTbar$-deformed system and the corresponding inverse temperatures $\left(\beta_u,\beta_v\right)$ of the auxiliary system are given by,
	\begin{equation}
		\begin{aligned}
			\beta_U=&\beta_u+2\mu\bar{\kappa}\bar{\mathcal{L}}_\mu \beta_v=\frac{\pi}{\sqrt{\mathcal{L}_\mu}}+2\mu\pi\bar{\kappa}\sqrt{\bar{\mathcal{L}}_\mu},\\ \beta_V=&\beta_v+2\mu\kappa\mathcal{L}_\mu\beta_u=\frac{\pi}{\sqrt{\bar{\mathcal{L}}_\mu}}+2\mu\pi\kappa\sqrt{\mathcal{L}_\mu},
		\end{aligned}
	\end{equation}
	where we have used \eqref{nodtem}. We can express the parameters $\left(\mathcal{L}_\mu,\bar{\mathcal{L}}_\mu\right)$ in terms of the inverse temperatures $\left(\beta_U,\beta_V\right)$,
	\begin{equation}\label{TMGparameters}
		\begin{aligned}
			\sqrt{\mathcal{L}_\mu}=&\frac{2\pi\beta_V}{\beta_U\beta_V+2\pi^2\left(\kappa-\bar{\kappa}\right)\mu+\sqrt{\left(\beta_U\beta_V+2\pi^2\left(\kappa-\bar{\kappa}\right)\mu\right)^2-8\pi^2\kappa\mu\beta_U\beta_V}},\\
			\sqrt{\bar{\mathcal{L}}_\mu}=&\frac{2\pi\beta_U}{\beta_U\beta_V-2\pi^2\left(\kappa-\bar{\kappa}\right)\mu+\sqrt{\left(\beta_U\beta_V-2\pi^2\left(\kappa-\bar{\kappa}\right)\mu\right)^2-8\pi^2\bar{\kappa}\mu\beta_U\beta_V}}.
		\end{aligned}
	\end{equation}
	
	Consider the single interval $A$ in the $\TTbar$-deformed system,
	\begin{equation}\label{interval A}
		A:\left(-\Delta U/2,-\Delta V/2\right)\rightarrow \left(\Delta U/2,\Delta V/2\right),
	\end{equation}
	which corresponds an interval in the auxiliary system, which is dual to the auxiliary Ba\~{n}ados geometry \eqref{auxiliary geometry}, according to the coordinate transformations \eqref{coordinatetrans},
	\begin{equation}\label{auxinterval}
		\tilde{A}:\left(-\Delta u/2,-\Delta v/2\right)\rightarrow \left(\Delta u/2,\Delta v/2\right),
	\end{equation}
	where the lengths $\left(\Delta u,\Delta v\right)$
	read,
	\begin{equation}\label{auxlength}
		\Delta u=\frac{\Delta U+2\mu\bar{\kappa}\bar{\mathcal{L}}_\mu \Delta V}{1-4\mu^2\kappa\bar{\kappa}\mathcal{L}_\mu\bar{\mathcal{L}}_\mu},\quad \Delta v=\frac{\Delta V+2\mu\kappa\mathcal{L}_\mu \Delta U}{1-4\mu^2\kappa\bar{\kappa}\mathcal{L}_\mu\bar{\mathcal{L}}_\mu}.
	\end{equation}
	
	Following our prescription, the $\TTbar$-deformed holographic entanglement entropy of the interval $A$ \eqref{interval A} in the $\TTbar$-deformed system should be equal to that of $\tilde{A}$ \eqref{auxinterval} in the auxiliary system up to the coordinate transformations \eqref{coordinatetrans}, i.e.
	\begin{equation}\label{HEETTbar}
		\begin{aligned}
			S_A^n=&S_{\tilde{A}}^n=\frac{1}{4G}\text{Length}\left(\mathcal{E}_{\tilde{A}}\right)=\frac{1}{4G}\log\left[\frac{\sinh(\sqrt{\mathcal{L}_\mu}\Delta u)\sinh(\sqrt{\bar{\mathcal{L}}_\mu}\Delta v)}{\sqrt{\mathcal{L}_\mu\bar{\mathcal{L}}_\mu}\:\epsilon^2}\right],\\
			S_A^a=&S_{\tilde{A}}^a=\frac{1}{4\lambda G}\int_{\mathcal{E}_{\tilde{A}}}d\tau\: \tilde{\mathbf{n}}\cdot\nabla\mathbf{n}=\frac{1}{4\lambda G}\log\left[\frac{\sinh(\sqrt{\mathcal{L}_\mu}\Delta u)\sqrt{\bar{\mathcal{L}}_\mu}}{\sinh(\sqrt{\bar{\mathcal{L}}_\mu}\Delta v)\sqrt{\mathcal{L}_\mu}}\right],
		\end{aligned}
	\end{equation}
	where $\left(\Delta u,\Delta v\right)$ and $\left(\mathcal{L}_\mu,\bar{\mathcal{L}}_\mu\right)$ are given by \eqref{auxlength} and \eqref{parelaTMG}, respectively, and $\mathcal{E}_{\tilde{A}}$ is the RT surface of the interval $\tilde{A}$ in the auxiliary system. Note that, our choice of the normal frame does not change under the coordinate transformation \eqref{coordinatetrans}, as we still choose the unit vector along the temporal direction of the auxiliary Ba\~{n}ados geometry \eqref{auxiliary geometry}, not that of the deformed geometry \eqref{deformed geometry}. The above result \eqref{HEETTbar} is non-perturbative in the finite deformation parameter $\mu$. In fact, once the deformed geometry \eqref{deformed geometry} is obtained, we may also calculate the entanglement entropy using CFT techniques, where the CFT is equipped with a metric that is given by the induced metric at the asymptotic boundary. We leave the detailed calculation to Appendix.\ref{Sec induced metric}. Furthermore, we discuss the violation of the boosted strong additivity of the entanglement entropy in Appendix.\ref{Sec stradd}, which indicates the non-local nature of the $\TTbar$ deformation.
	
	In order to compare the non-perturbative result \eqref{HEETTbar} with the perturbative result evaluated from the field theory side, we expand it to the first order in the deformation parameter $\mu$, and the leading order correction to the holographic entanglement entropy with gravitational anomalies is then given as
	\begin{equation}
		\begin{aligned}
			\delta S_A=\delta S_A^n+\delta S_A^a
			=&\frac{c^{}_Lc^{}_R\pi\mu}{36\beta_U^2\beta_V^2}\Big(-2\beta_U\beta_V\\
			&+\pi\left(\beta_V\Delta U+\beta_U\Delta V\right)\left(\coth\left(\frac{\pi \Delta U}{\beta_U}\right)+\coth\left(\frac{\pi \Delta V}{\beta_V}\right)\right)\Big),
		\end{aligned}
	\end{equation}
	where we recover the deformation parameter $\mu$, and have used the Brown-Henneaux relation \eqref{twocentreal} in TMG. In fact, as pointed out in \cite{Chen:2018eqk}, the bulk rotating BTZ geometry is dual to the  CFT only in the high temperature limit, i.e., $\beta_U,\beta_V \rightarrow 0$ while keeping $\beta_U/\beta_V$ an order $1$ constant (i.e. $\beta\rightarrow 0$ keeping $\Omega$ finite). This implies that the first term in the above expression can be neglected compared with the second term, so that this expression reduces to the perturbative result \eqref {EE-corrections} evaluated from the field theory side in the high temperature limit. Note that, in the leading order in $\mu$, the temperatures $(\beta_{U},\beta_{V})$ reduce to the undeformed values $\beta_\pm$.
	
	In particular, we can consider two interesting limits, the non-anomalous limit $\lambda\rightarrow \infty$ (i.e. $c^{}_L=c^{}_R$) and the chiral limit $\lambda\rightarrow \pm 1$ (i.e. $c^{}_R=0$ or $c^{}_L=0$). At first, we consider the case that the interval $A$ is purely spacelike with length $l$, and take the non-anomalous limit $\lambda\rightarrow \infty$, our result \eqref{HEETTbar} indeed reduces to the following result reported in \cite{Banerjee:2024wtl},  
	\begin{equation}\label{HEEnonga}
		S_A=\frac{1}{4G}\log\left(8\pi^2\mu^2\frac{\cosh\left(\frac{\pi l\left(\beta_U+\beta_V\right)}{\beta_U\beta_V\sqrt{1-\frac{8\pi^2\mu}{\beta_U\beta_V}}}\right)-\cosh\left(\frac{\pi l\left(\beta_U-\beta_V\right)}{\beta_U\beta_V}\right)}{\epsilon^2\beta_U\beta_V\left(1-\sqrt{1-\frac{8\pi^2\mu}{\beta_U\beta_V}}\right)^2}\right).
	\end{equation} 
	The reality of the entanglement entropy imposes non-trivial lower and upper bounds for the deformation parameter $\mu$ \cite{Banerjee:2024wtl},
	\begin{equation}\label{boundnon-ga}
		-\frac{\beta^2\left(1-\Omega^2\right)^2}{8\pi^2\Omega^2}<\mu<\frac{\beta^2\left(1-\Omega^2\right)}{8\pi^2},
	\end{equation}
	where the upper bound known as the Hagedorn bound\footnote{The Hagedorn bound, in fact, arises from the singularity of the torus partition function \cite{Cavaglia:2016oda,Dubovsky:2012wk}, as demonstrated in subsection \ref{Hagedorn} in the context of gravitational anomaly.} and comes from the zero point of the expression under the square roots \cite{McGough:2016lol}; hence the entanglement entropy will diverge here. The lower bound comes from the zero-point of the logarithm, where the geodesic connecting the endpoints of the subsystem $A$ becomes null. 
	
	Similarly, we can also consider the chiral limits $\lambda=\pm 1$, whence,
	\begin{equation}
		S_A=\begin{cases} 
			\frac{c^{}_L}{6} \log\left[\frac{\beta_U}{\pi\epsilon}\sinh\left(\frac{\pi\Delta U}{\beta_U}\right)\right] & \text{when } \lambda\rightarrow 1, \\
			\frac{c^{}_R}{6} \log\left[\frac{\beta_V}{\pi\epsilon}\sinh\left(\frac{\pi\Delta V}{\beta_V}\right)\right] & \text{when } \lambda\rightarrow -1.
		\end{cases}
	\end{equation}
	Interestingly, both of these expressions do not depend on the deformation parameter $\mu$, which implies that the $\TTbar$ deformation does not change the holographic entanglement entropy of the chiral CFT, although it still introduces the field-dependent coordinate transformation for one certain sector.

	\subsection{Bounds on the deformation parameter}\label{sec:Bounds-on-mu}
	
\begin{figure}
		\centering
		\includegraphics[width=0.7\linewidth]{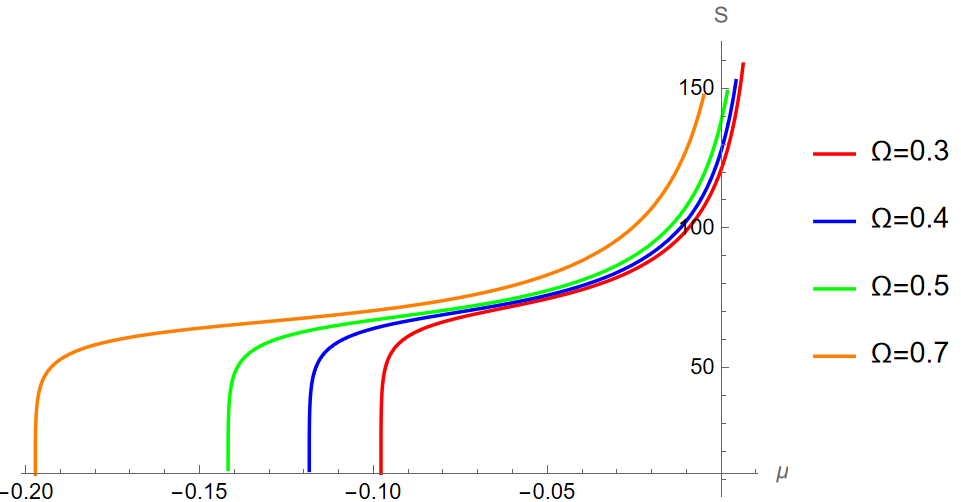}
		\caption{Plot of the $\TTbar$-deformed holographic entanglement entropy with respect to the deformation parameter $\mu$. We have set $l=2,\beta=1,\epsilon=0.01,G=1/8\pi$ and $\lambda=2$.}
		\label{fig:sn1}
	\end{figure}

\begin{figure}
		\centering
		\includegraphics[width=0.7\linewidth]{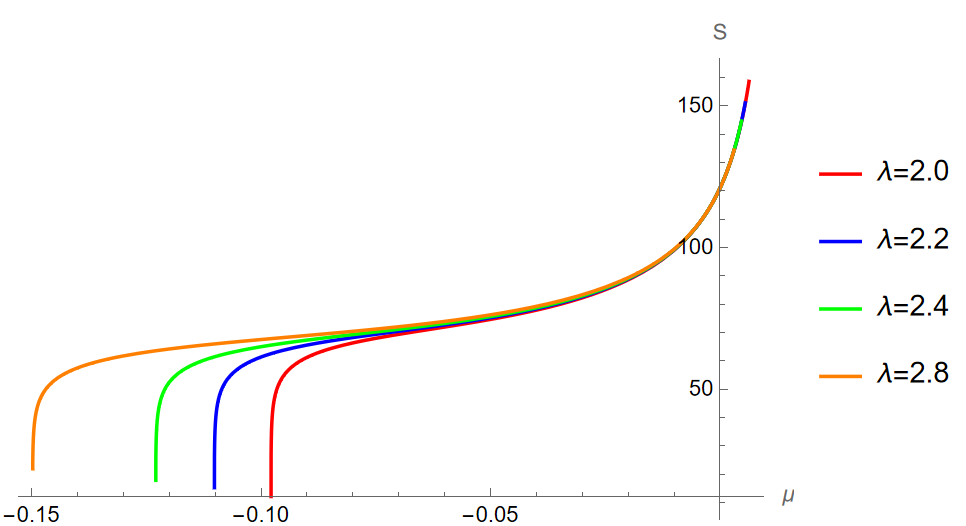}
		\caption{Plot of the $\TTbar$-deformed holographic entanglement entropy with respect to the coupling constant $\lambda$. We have set $l=2,\beta=1,\epsilon=0.01,G=1/8\pi$ and $\Omega=0.3$.}
		\label{fig:sn2}
	\end{figure}

	In this subsection, we impose the reality condition on the holographic entanglement entropy \eqref{HEETTbar} of the anomalous CFT with $\TTbar$ deformation to study the physical bounds on the deformation parameter $\mu$. We consider a purely spacelike interval of length $l$. The reality of the holographic entanglement entropy \eqref{HEETTbar} requires,
	\begin{equation}
		\begin{aligned}
			\mathcal{L}_\mu&>0,\quad \bar{\mathcal{L}}_\mu>0,\quad 
			\Delta u>0,\quad \Delta v>0.
		\end{aligned}
	\end{equation}
	where $\left(\mathcal{L}_\mu,\bar{\mathcal{L}}_\mu\right)$ are given by \eqref{TMGparameters}, and $\left(\Delta u,\Delta v\right)$ are given by \eqref{auxlength} with $\Delta U=\Delta V=l$. Firstly, the zero points of the expression under the square roots in $\mathcal{L}_\mu$ (as well as $\bar{\mathcal{L}}_\mu$) are,
	\begin{equation}
		\mu_1=\frac{\beta^2\left(1-\Omega^2\right)}{4\pi^2}\lambda\left(\lambda-\sqrt{\lambda^2-1}\right),\quad \mu_2=\frac{\beta^2\left(1-\Omega^2\right)}{4\pi^2}\lambda\left(\lambda+\sqrt{\lambda^2-1}\right),
	\end{equation}
	which are both larger than $0$. In fact, these are identical to the generalized Hagedorn bound \eqref{T-max} for the anomalous CFT, and we will see that they are indeed the upper bounds of the deformation parameter $\mu$. These will reduce to the upper bound in \eqref{boundnon-ga} in the non-anomalous limits $\lambda\rightarrow +\infty$ and $\lambda\rightarrow -\infty$, respectively. Since their relative strength depends on whether $\lambda>1$ or $\lambda<-1$, we divide into two cases to consider in the following. We first consider the case $\lambda>1$. In this case, $\mu_1<\mu_2$, and the condition $\mathcal{L}_\mu,\bar{\mathcal{L}}_\mu>0$ further requires,
	\begin{equation}\label{mumu1}
		\mu<\mu_1.
	\end{equation}
	Second, we impose the condition $\Delta u,\Delta v>0$, which implies,
	\begin{align}
		1+2\mu\kappa\mathcal{L}_\mu&>0,\label{1+2mukappa}\\
		1+2\mu\bar{\kappa} \bar{\mathcal{L}}_\mu&>0,\label{1+2mukappabar}\\ 1-4\mu^2\kappa\bar{\kappa}\mathcal{L}_\mu\bar{\mathcal{L}}_\mu&>0.\label{1-4}
	\end{align}
	Note that \eqref{1-4} is always satisfied in the range \eqref{mumu1}, hence we focus on \eqref{1+2mukappa} and \eqref{1+2mukappabar} in the following. There is a unique zero point of \eqref{1+2mukappa},
	\begin{equation}
		1+2\mu\kappa\mathcal{L}_\mu=0\quad \text{at}\quad \mu=\mu_{c,1}=-\frac{\beta^2\lambda}{4\pi^2}\left(1+\lambda \Omega^2+\Omega\sqrt{\left(1+\lambda\right)\left(2+\left(\lambda-1\right)\Omega^2\right)}\right),
	\end{equation}
	which keeps the entropy real-valued in the range \eqref{mumu1}. Hence \eqref{1+2mukappa} is always satisfied when
	\begin{equation}\label{range1}
		\mu_{c,1}<\mu<\mu_1.
	\end{equation} 
	On the other hand, there are two zero points of \eqref{1+2mukappabar},
	\begin{equation}
		\begin{aligned}
			1+2\mu\bar{\kappa}\bar{\mathcal{L}}_\mu=0\quad \text{at}\quad \mu=&\mu_{c,2}=\frac{\beta^2\lambda}{4\pi^2}\left(1-\lambda\Omega^2-\Omega\sqrt{\left(\lambda-1\right)\left(-2+\left(\lambda+1\right)\Omega^2\right)}\right),\\ \text{and}\quad \mu=&\mu_{c,3}=\frac{\beta^2\lambda}{4\pi^2}\left(1-\lambda\Omega^2+\Omega\sqrt{\left(\lambda-1\right)\left(-2+\left(\lambda+1\right)\Omega^2\right)}\right).
		\end{aligned}
	\end{equation}
	Note that there is a critical point $\lambda_{c}$ for those two zero points,
	\begin{equation}
		\lambda_{c}=-1+\frac{2}{\Omega^2}.\quad \mu_{c,2},\:\mu_{c,3}\notin\text{Reals, when } \lambda<\lambda_{c},\quad
		\mu_{c,2},\:\mu_{c,3}\in\text{Reals, when } \lambda>\lambda_{c}.
	\end{equation}
	This implies
	\begin{equation}\label{range2}
		1+2\mu\bar{\kappa}\bar{\mathcal{L}}_\mu 
		\begin{cases} 
			>0 & \text{when } 1<\lambda<\lambda_c,\:\mu<\mu_1, \\
			<0 & \text{when } \lambda>\lambda_c,\:\mu_{c,1}<\mu<\mu_{c,2}\:\text{or}\:\mu_{c,3} <\mu<\mu_1.
		\end{cases}
	\end{equation}
	In summary, by combining \eqref{range1} with \eqref{range2}, we can obtain the bound behavior of the deformation parameter $\mu$ for the anomalous CFT,
	\begin{equation}
		\begin{aligned}
			\mu_{c,1}<	\mu<\mu_1,\quad \text{when}&\quad 1<\lambda<\lambda_c\\
			\mu_{c,1}<	\mu<\mu_{c,2},\quad \text{or}\quad \mu_{c,3}<	\mu<\mu_1 \text{ when}&\quad \lambda>\lambda_c.
		\end{aligned}
	\end{equation}
	Fig.\ref{fig:sn1} and Fig.\ref{fig:sn2} show the $\TTbar$-deformed holographic entanglement entropy with gravitational anomalies as a function of the deformation parameter $\mu$, for various values of $\Omega$ and $\lambda$, respectively.
	
	
	At last, let us comment the non-anomalous limit of these critical points,
	\begin{equation}
		\mu_1\rightarrow \frac{\beta^2\left(1-\Omega^2\right)}{8\pi^2},\quad \mu_{c,1},\:\mu_{c,2}\rightarrow -\infty,\quad \mu_{c,3} \rightarrow  -\frac{\beta^2\left(1-\Omega^2\right)^2}{8\pi^2\Omega^2}\: \text{when}\: \lambda\rightarrow+\infty.
	\end{equation} 
	Therefore, in this limit, the bound behavior becomes,
	\begin{equation}
		-\frac{\beta^2\left(1-\Omega\right)^2}{8\pi^2\Omega^2}<\mu<\frac{\beta^2\left(1-\Omega^2\right)}{8\pi^2},
	\end{equation}
	which is the same as \eqref{boundnon-ga} of non-anomalous CFT as we expected. Similarly, one may easily find the bound behavior of the deformation parameter $\mu$ when $\lambda<-1$,
	\begin{equation}
		\mu_{c,3}<\mu<\mu_2,
	\end{equation}
	which will also reduce to \eqref{boundnon-ga} in the non-anomalous limit $\lambda\rightarrow -\infty$.

	\subsection{Entanglement wedge cross section}

	In this subsection, we continue to calculate the EWCS for various bipartite mixed states in the deformed bulk geometry following the earlier prescription. 
	
	\subsubsection{Non-adjacent case}
	
	Consider two non-adjacent covariant intervals $A$ and $B$ in the $\TTbar$-deformed system,
	\begin{equation}\label{twoinABnon}
		A:\left(U_1,V_1\right)\rightarrow \left(U_2,V_2\right),\quad B:\left(U_3,V_3\right)\rightarrow \left(U_4,V_4\right),
	\end{equation}
	which also correspond two covariant intervals $\tilde{A}$ and $\tilde{B}$ at the boundary of the auxiliary Ba\~{n}ados geometry \eqref{auxiliary geometry},
	\begin{equation}\label{twoinABnonaux}
		\tilde{A}:\left(u_1,v_1\right)\rightarrow \left(u_2,v_2\right),\quad \tilde{B}:\left(u_3,v_3\right)\rightarrow \left(u_4,v_4\right),
	\end{equation}
	where the relations between $\left(u_i,v_i\right)$ and $\left(U_i,V_i\right)$ are given by the coordinate transformation \eqref{coordinatetrans}. According to our prescription, the EWCS with gravitational anomalies between $A$ and $B$ in the $\TTbar$-deformed system should be equal to that between $\tilde{A}$ and $\tilde{B}$ in the auxiliary system up to the coordinate transformations \eqref{coordinatetrans}. Therefore, it can be obtained by applying \eqref{coordinatetrans} into \eqref{nonEWCS}, i.e.
	\begin{equation}\label{deEWCSnon}
		\begin{aligned}
			E_W^n\left(A:B\right)=&\frac{1}{8G}\text{Length}\left(\Sigma_{AB}\right)=\frac{1}{8G}\log\left[\frac{\left(1+\sqrt{\mathcal{Q}}\right)\left(1+\sqrt{\bar{\mathcal{Q}}}\right)}{\left(1-\sqrt{\mathcal{Q}}\right)\left(1-\sqrt{\bar{\mathcal{Q}}}\right)}\right],\\
			E_W^a\left(A:B\right)=&\frac{1}{8\lambda G}\int_{\Sigma_{AB}}d\tau\: \tilde{\mathbf{n}}\cdot\nabla\mathbf{n} =\frac{1}{8\lambda G}\log\left[\frac{\left(1+\sqrt{\mathcal{Q}}\right)\left(1-\sqrt{\bar{\mathcal{Q}}}\right)}{\left(1-\sqrt{\mathcal{Q}}\right)\left(1+\sqrt{\bar{\mathcal{Q}}}\right)}\right],
		\end{aligned}
	\end{equation}
	where $\Sigma_{AB}$ is the saddle geodesic connecting the two pieces of the RT surfaces of $A\cup B$ in the deformed bulk geometry \eqref{deformed geometry}; see Fig.\ref{fig:non-adj} for an illustration. The parameters $\left(\mathcal{Q},\bar{\mathcal{Q}}\right)$
	are given as,
	\begin{equation}
		\mathcal{Q}=\frac{\sinh\left(\sqrt{\mathcal{L}_\mu} u_{12}\right)\sinh\left(\sqrt{\mathcal{L}_\mu} u_{34}\right)}{\sinh\left(\sqrt{\mathcal{L}_\mu} u_{13}\right)\sinh\left(\sqrt{\mathcal{L}_\mu} u_{24}\right)},\quad \bar{\mathcal{Q}}=\frac{\sinh\left(\sqrt{\bar{\mathcal{L}}_\mu} v_{12}\right)\sinh\left(\sqrt{\bar{\mathcal{L}}_\mu} v_{34}\right)}{\sinh\left(\sqrt{\bar{\mathcal{L}}_\mu} v_{13}\right)\sinh\left(\sqrt{\bar{\mathcal{L}}_\mu} v_{24}\right)},
	\end{equation}
	and $\left(\mathcal{L}_\mu,\bar{\mathcal{L}}_\mu\right)$ and $\left(u_i,v_j\right)$ in the above formula are given by \eqref{TMGparameters} and \eqref{coordinatetrans}, respectively. As earlier, to compare this result with the perturbative result of the reflected entropy \eqref{SR-disj-corrections}, we expand \eqref{deEWCSnon} to the leading order in the deformation parameter $\mu$ as follows, 
	\begin{equation}
		\begin{aligned}
			\delta E_W\left(A:B\right)=&\delta E_W^n\left(A:B\right)+\delta E_W^a\left(A:B\right)\\
			=&\frac{c^{}_Lc^{}_R\pi^2\mu}{72\beta_U^2\beta_V^2}\left[\sqrt{\eta}\left(\mathcal{P}_{12}+\mathcal{P}_{34}-\mathcal{P}_{14}-\mathcal{P}_{23}\right)+\sqrt{\bar{\eta}}\left(\bar{\mathcal{P}}_{12}+\bar{\mathcal{P}}_{34}-\bar{\mathcal{P}}_{14}-\bar{\mathcal{P}}_{23}\right)\right],
		\end{aligned}
	\end{equation}
	where $\left(\eta,\bar{\eta}\right)$ and $\left(\mathcal{P}_{ij},\bar{\mathcal{P}}_{ij}\right)$ are given by,
	\begin{equation}
		\begin{aligned}
			\eta=&\frac{\sinh\left(\frac{\pi U_{12}}{\beta_U}\right)\sinh\left(\frac{\pi U_{34}}{\beta_U}\right)}{\sinh\left(\frac{\pi U_{13}}{\beta_U}\right)\sinh\left(\frac{\pi U_{24}}{\beta_U}\right)},\quad \bar{\eta}=\frac{\sinh\left(\frac{\pi V_{12}}{\beta_V}\right)\sinh\left(\frac{\pi V_{34}}{\beta_V}\right)}{\sinh\left(\frac{\pi V_{13}}{\beta_V}\right)\sinh\left(\frac{\pi V_{24}}{\beta_V}\right)},\\
			\mathcal{P}_{ij}=&\left(\beta_VU_{ij}+\beta_UV_{ij}\right)\coth\left(\frac{\pi U_{ij}}{\beta_U}\right),\quad \bar{\mathcal{P}}_{ij}=\left(\beta_VU_{ij}+\beta_UV_{ij}\right)\coth\left(\frac{\pi V_{ij}}{\beta_V}\right).
		\end{aligned}
	\end{equation}
	One can easily check that this result exactly matches one half of the reflected entropy \eqref{SR-disj-corrections}, even without taking the high-temperature limit here.

	\subsubsection{Adjacent case}
	
	Next, we consider two adjacent covariant intervals $A$ and $B$ in the $\TTbar$-deformed system,
	\begin{equation}
		A:\left(U_1,V_1\right)\rightarrow \left(U_2,V_2\right),\quad B:\left(U_2,V_2\right)\rightarrow \left(U_3,V_3\right),
	\end{equation}
	which correspond the following intervals at the boundary of the auxiliary Ba\~{n}ados geometry \eqref{auxiliary geometry},
	\begin{equation}
		\tilde{A}:\left(u_1,v_1\right)\rightarrow \left(u_2,v_2\right),\quad \tilde{B}:\left(u_2,v_2\right)\rightarrow \left(u_3,v_3\right).
	\end{equation}
	The two terms of the EWCS with gravitational anomalies between $A$ and $B$ in the $\TTbar$-deformed system, which represent the contributions from the Einstein-Hilbert term and the CS term, are given by applying \eqref{coordinatetrans} to \eqref{adjEWCS}:
	\begin{equation}
		\begin{aligned}
			E_W^n\left(A,B\right)=&\frac{1}{8G}\log\left[\frac{2}{\sqrt{\mathcal{L}_\mu} \epsilon}\frac{\sinh(\sqrt{\mathcal{L}_\mu} u_{12})\sinh(\sqrt{\mathcal{L}_\mu} u_{23})}{\sinh(\sqrt{\mathcal{L}_\mu} u_{13})}\right]\\
			+&\frac{1}{8G}\log\left[\frac{2}{\sqrt{\bar{\mathcal{L}}_\mu} \epsilon}\frac{\sinh(\sqrt{\bar{\mathcal{L}}_\mu} v_{12})\sinh(\sqrt{\bar{\mathcal{L}}_\mu} v_{23})}{\sinh(\sqrt{\bar{\mathcal{L}}_\mu} v_{13})}\right],\\
			E_W^a\left(A,B\right)=&\frac{1}{8\lambda G}\log\left[\frac{2}{\sqrt{\mathcal{L}_\mu} \epsilon}\frac{\sinh(\sqrt{\mathcal{L}_\mu} u_{12})\sinh(\sqrt{\mathcal{L}_\mu} u_{23})}{\sinh(\sqrt{\mathcal{L}_\mu} u_{13})}\right]\\
			-&\frac{1}{8\lambda G}\log\left[\frac{2}{\sqrt{\bar{\mathcal{L}}_\mu} \epsilon}\frac{\sinh(\sqrt{\bar{\mathcal{L}}_\mu} v_{12})\sinh(\sqrt{\bar{\mathcal{L}}_\mu} v_{23})}{\sinh(\sqrt{\bar{\mathcal{L}}_\mu} v_{13})}\right],
		\end{aligned}
	\end{equation}
	where $\left(\mathcal{L}_\mu,\bar{\mathcal{L}}_\mu\right)$ and $\left(u_i,v_j\right)$ in the above formula are given by \eqref{TMGparameters} and \eqref{coordinatetrans}, respectively. See Fig.\ref{fig:non-adj} for an illustration. To compare with the field theoretic computations, we again expand the holographic result in the deformation parameter $\mu$, and the leading order correction is,
	\begin{equation}
		\begin{aligned}
			\delta E_W\left(A:B\right)=&\delta E_W^n\left(A:B\right)+\delta E_W^a\left(A:B\right)\\
			=&\frac{\mu \pi c^{}_Lc^{}_R}{72\beta_U^2\beta_V^2}\left(-2\beta_U\beta_V+\pi\left(\mathbb{P}_{12}+\mathbb{P}_{23}-\mathbb{P}_{13}\right)\right),
		\end{aligned}
	\end{equation}
	where $\mathbb{P}_{ij}$ is defined as,
	\begin{equation}
		\mathbb{P}_{ij}=\left(\beta_VU_{ij}+\beta_U V_{ij}\right)\left(\coth\left(\frac{\pi U_{ij}}{\beta_U}\right)+\coth\left(\frac{\pi V_{ij}}{\beta_V}\right)\right).
	\end{equation}
	One can easily check that the above result matches the field theoretic perturbative result \eqref{SR-adj-correction} evaluated using the conformal perturbation theory when taking the high-temperature limit.

	\subsection{Holographic balanced partial entanglement entropy}
	
    The balanced partial entanglement entropy (BPE) \cite{Wen:2021qgx,Camargo:2022mme} is  a quantity defined on quantum systems without holography, and is proposed to characterize the mixed-state entanglement and is conjectured to be dual to the EWCS in holography. In two-dimensional quantum field theories, the BPE can be expressed as a special linear combination of the entanglement entropy for certain single intervals. Unlike the reflected entropy, which is defined in the canonical purification, the BPE can be defined in a generic purification of $\rho_{A\cup B}$, including the boundary state in which $A$ and $B$ are embedded.
	
	In this subsection, in the context of the correspondence between the $\TTbar$-deformed anomalous CFT$_2$ and the TMG with the mixed boundary conditions, we argue that the BPE continues to reproduce the EWCS with the CS correction in the $\TTbar$-deformed gravity dual. We will use the holographic entanglement entropy \eqref{HEETTbar} for single intervals to construct the BPE and match it to the EWCS \eqref{deEWCSnon} with CS correction. This matching goes beyond the perturbative level. One can also conduct a non-perturbative computation for the reflected entropy with the indication from the bulk dynamical coordinate transformations following the method sketched in Appendix.\ref{Sec induced metric}, and check its matching with the EWCS.
	
	\subsubsection{The definition of BPE}
	
\begin{figure}
		\centering
		\includegraphics[width=0.7\linewidth]{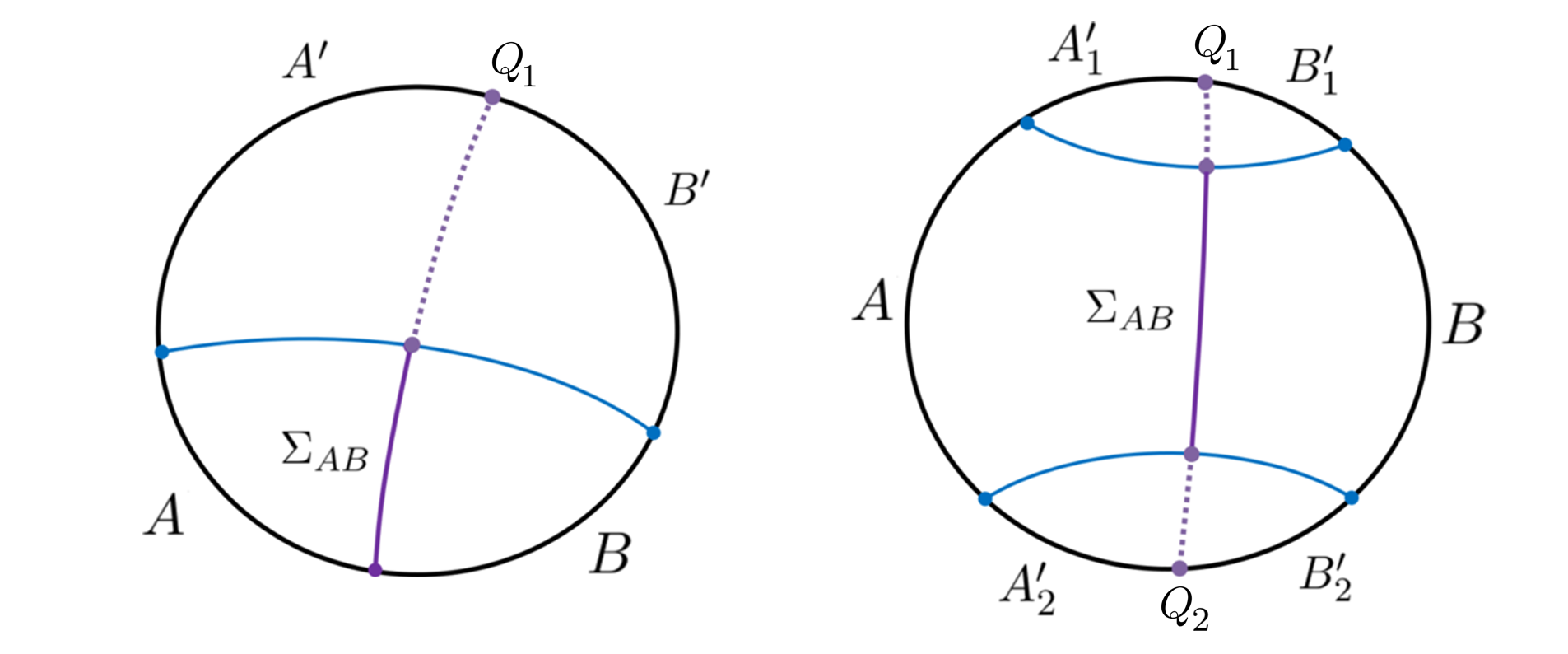}
		\caption{The figures are extracted from \cite{Wen:2021qgx}. The region $A'B'$ serves as an auxiliary system that purifies the mixed state $\rho_{AB}$. The solid blue lines represent the RT surfaces for $A\cup B$, the solid purple lines represent the EWCS $\Sigma_{AB}$ for the entanglement wedge of $A\cup B$, and the dashed purple line represent the extension of $\Sigma_{AB}$, which are the geodesics that anchor on the balanced partition points $Q_i$. }
		\label{fig:bpepicture}
	\end{figure}

	The balanced partial entanglement entropy (BPE) is a special type of the partial entanglement entropy (PEE) that satisfies the so-called balanced conditions. The PEE $s_A\left(A_i\right)$ is a local measure of how much a subset $A_i \subset A$ contributes to the entanglement entropy of a subsystem $A$ \cite{Vidal:2014aal,Wen:2018whg,Wen:2019iyq}, hence can be expressed as a integral of the entanglement contour function $s_A\left(x\right)$ that represents the density of the entanglement entropy,
	\begin{equation}
		s_A\left(A_i\right)=\int_{A_i} dx\: s_A\left(x\right).
	\end{equation} 
	The PEE satisfies several physical requirements including additivity, normalization, and so on \cite{Vidal:2014aal,Wen:2018whg,Wen:2019iyq}. Up to now, there have been many methods to construct the PEE, including field theoretical and holographic approaches \cite{Wen:2018whg,Wen:2019iyq,Han:2019scu}. The most convenient one is the so called additive linear combination (ALC) proposal \cite{Wen:2018whg,Wen:2019iyq} which applies to the two dimensional quantum field theories and satisfies all the physical requirements of the PEE. Consider an interval $A$ which can be partitioned into three sub-intervals $A_1\cup A_2\cup A_3$ where $A_2$ is the middle one. The proposal then claims that
	\begin{equation}\label{ALC proposal}
		s_A\left(A_2\right)=\frac{1}{2}\left(S_{A_1A_2}+S_{A_2A_3}-S_{A_1}-S_{A_2}\right).
	\end{equation}
	Now, consider a bipartite system $\mathcal{H}_A\otimes \mathcal{H}_B$ with the density matrix $\rho_{AB}$. One can introduce an auxiliary system $A'B'$ to purify it such that the whole system $ABA'B'$ is in a pure state $\ket{\psi}$, and satisfies
	\begin{equation}
		\text{Tr}_{A'B'}\ket{\psi}\bra{\psi}=\rho_{AB},\quad \ket{\psi}\in \mathcal{H}_A\otimes \mathcal{H}_B\otimes \mathcal{H}_{A'}\otimes \mathcal{H}_{B'}.
	\end{equation}
	Then the BPE between $A$ and $B$ is then defined as \cite{Wen:2021qgx,Camargo:2022mme},
	\begin{equation}\label{BPEde}
		\text{BPE}\left(A,B\right) =s_{AA'}\left(A\right)|_{\text{balanced}}
	\end{equation}
	where the subscript ``balanced'' means that the auxiliary system $A'B'$ satisfies the balance conditions, which are listed in the following,
	\begin{itemize}
		\item  When $A$ and $B$ are adjacent, the balance condition is
		\begin{equation}
			s_{AA'}\left(A\right)=s_{BB'}\left(B\right).
		\end{equation}
		\item When $A$ and $B$ are non-adjacent, we have two balance conditions,
		\begin{equation}
			s_{AA'}\left(A'_1\right)=s_{BB'}\left(B_1'\right),\quad s_{AA'}\left(A\right)=s_{BB'}\left(B\right),
		\end{equation}
		where $A_1'$ are $A_2'$ are two sub-intervals of $A'$ sandwiching $A$, similarly for $B_1'$ and $B_2'$, see Fig.\ref{fig:bpepicture} for an illustration.
	\end{itemize}
	
	Solving the balance conditions, one can determine how to divide the purification system $A'B'$ and identify the balanced points $Q_1$ and $Q_2$ for $A'B'$. It was shown that the BPE \eqref{BPEde} exactly captures the same type of mixed state correlations as the reflected entropy \cite{Wen:2021qgx,Camargo:2022mme,Wen:2022jxr,Basu:2023wmv}, and reproduces the length of the EWCS $\Sigma_{AB}$. Furthermore, the balanced points $Q_1$ and $Q_2$ that satisfy the balance conditions correspond precisely the locations where the extension of the EWCS anchors on the asymptotic boundary.
	
	\subsubsection{$\TTbar$-deformed BPE with gravitational anomalies}
	
\begin{figure}
		\centering
		\includegraphics[width=0.7\linewidth]{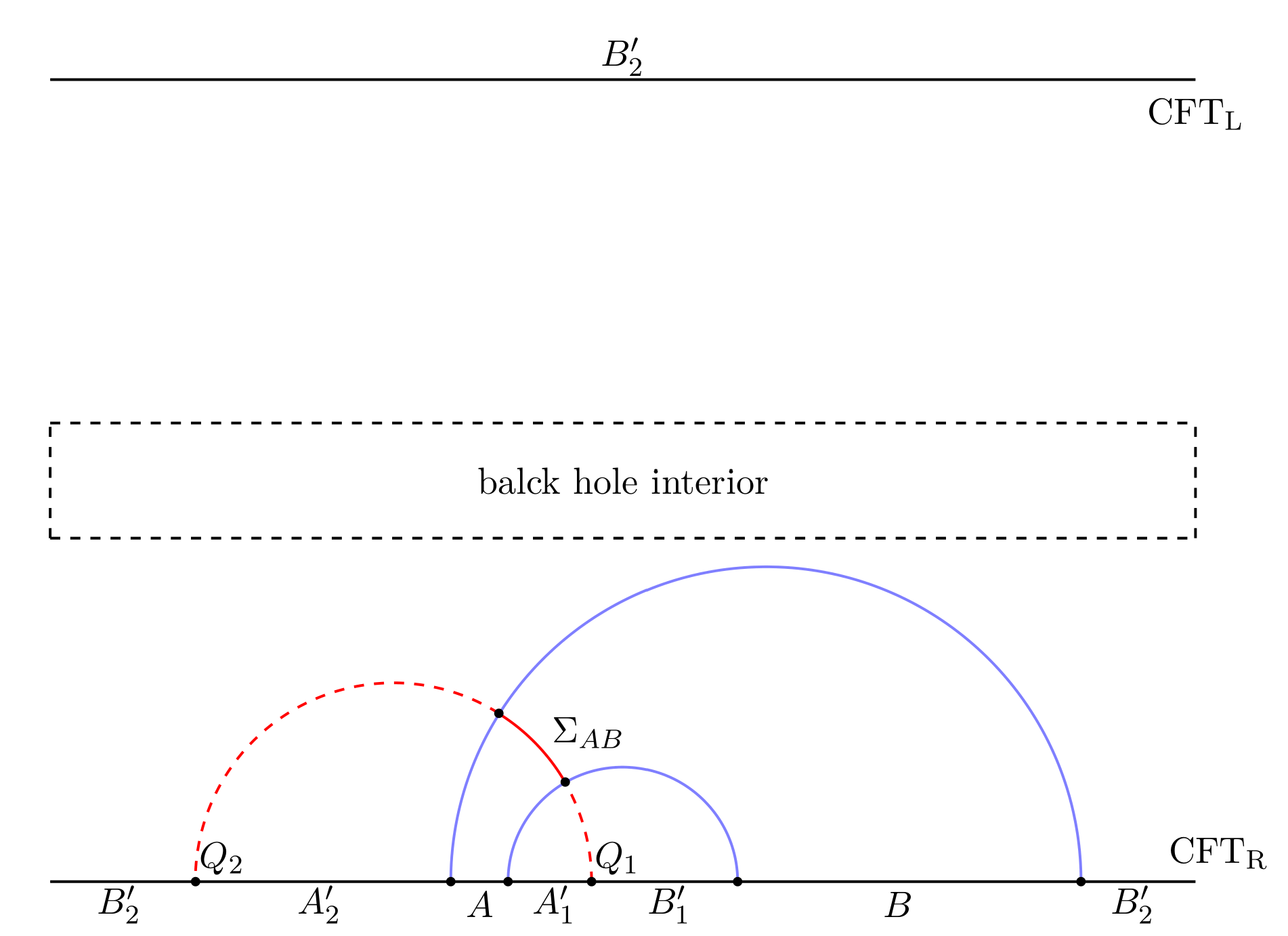}
		\caption{We use the thermofield double (TFD) state as the purification for the boundary state. The red solid line represent the EWCS $\Sigma_{AB}$, whose extension is a geodesic that anchors on the boundary at exactly the balanced points $Q_1$ and $Q_2$. Here, both of the balanced partition points $Q_1$ and $Q_2$ are located in CFT$_R$, and the CFT$_L$ belongs to $B_2'$.}
		\label{fig:bpepart}
	\end{figure}

	As we argued in the last subsection, the EWCS of the two intervals $A$ and $B$ given by \eqref{twoinABnon} in the $\TTbar$-deformed system is equal to that of the corresponding intervals $\tilde{A}$ and $\tilde{B}$ in \eqref{twoinABnonaux} in the auxiliary system. In fact, this argument can be generalized directly to the BPE. Moreover, it has been shown that the BPE exactly matches the EWCS in the auxiliary system (which is a CFT without the $\TTbar$ deformation) in \cite{Wen:2022jxr}, which implies that the correspondence between the BPE and the EWCS also holds in the $\TTbar$-deformed system. More explicitly, let us consider the case of two non-adjacent intervals $A$ and $B$ \eqref{twoinABnon} as an example. Unlike the configurations studied previously, the $\TTbar$-deformed boundary state we considered is a thermal state, which necessitates an auxiliary system outside the boundary to make a pure state. In order to compute the $\text{BPE}\left(A,B\right)$, we consider the thermofield double (TFD) state, whose gravity dual is an eternal black hole. We denote the CFT that contains $A$ and $B$ as CFT$_{\text{R}}$, and the purifying system as CFT$_{\text{L}}$. Furthermore, we only consider the configurations, where the balanced points $Q_i$, obtained by solving the balance conditions, locate in CFT$_R$\footnote{This corresponds to the cases where the extension of the EWCS does not cross the horizon to connect the other asymptotic boundary, where CFT$_L$ is located.}. Consequently, the purifying system CFT$_L$ does not need to be partitioned and belongs to either $B_1'$ or $B_2'$. See Fig.\ref{fig:bpepart} for an illustration.

	For the CFT with gravitational anomalies, the entanglement entropy can be divided into two types, the normal contribution $S_A^n$ from the Einstein-Hilbert action and the anomalous contribution $S_A^a$ from the CS term. Accordingly, the BPE (as well as PEE) can also be divided into these two types,
	\begin{equation}\label{twopartsBPE}
		\text{BPE}^n\left(A,B\right)=s_{AA'}^n\left(A\right)|_{\text{balanced}},\quad \text{BPE}^a\left(A,B\right)=s_{AA'}^a\left(A\right)|_{\text{balanced}}.
	\end{equation}  
	Note that, the balance conditions should be imposed to both the normal and the anomalous contributions independently \cite{Wen:2022jxr}. For the non-adjacent case, there are four balanced conditions,
	\begin{equation}\label{fourbalanced}
		\begin{aligned}
			s_{AA'}^n\left(A_1'\right)&=s_{BB'}^n\left(B_1'\right),\quad s_{AA'}^a\left(A_1'\right)=s_{BB'}^a\left(B_1'\right),\\ s_{AA'}^n\left(A\right)&=s_{BB'}^n\left(B\right),\quad s_{AA'}^a\left(A\right)=s_{BB'}^a\left(B\right),
		\end{aligned}
	\end{equation}
which are enough to determine the coordinates of the two balanced points $Q_1$ and $Q_2$. Using the ALC proposal \eqref{ALC proposal}, the above balanced conditions become,
	\begin{equation}
		\begin{aligned}
			S_{A_1'}^n-S_{B_1'}^n=&S_{AA_2'}^n-S_{BB_2'}^n,\quad S_{A_2'}^n-S_{B_2'}^n=S_{AA_1'}^n-S_{BB_1'}^n,\\
			S_{A_1'}^a-S_{B_1'}^a=&S_{AA_2'}^a-S_{BB_2'}^a,\quad ~~S_{A_2'}^a-S_{B_2'}^a=S_{AA_1'}^a-S_{BB_1'}^a.
		\end{aligned}
	\end{equation}
	Then we plug \eqref{HEETTbar} into above equations to solve the coordinates of the balanced points $Q_1$ and $Q_2$, which are given by,
	\begin{equation}
		\begin{aligned}
			u_{Q_1}=&\frac{\beta_u}{2\pi} \log \left(\frac{\sqrt{\mathcal{Y}_{12} \mathcal{Y}_{13} \mathcal{Y}_{24}\mathcal{Y}_{34}}+\mathcal{X}_{14}-\mathcal{X}_{23}}{\mathcal{Y}_{12}-\mathcal{Y}_{34}}\right),\\ v_{Q_1}=&\frac{\beta_v}{2\pi} \log \left(\frac{\sqrt{\tilde{\mathcal{Y}}_{12} \tilde{\mathcal{Y}}_{13} \tilde{\mathcal{Y}}_{24}\tilde{\mathcal{Y}}_{34}}+\tilde{\mathcal{X}}_{14}-\tilde{\mathcal{X}}_{23}}{\tilde{\mathcal{Y}}_{12}-\tilde{\mathcal{Y}}_{34}}\right),\\ u_{Q_2}=&\frac{\beta_u}{2\pi} \log \left(\frac{-\sqrt{\mathcal{Y}_{12} \mathcal{Y}_{13} \mathcal{Y}_{24}\mathcal{Y}_{34}}+\mathcal{X}_{14}-\mathcal{X}_{23}}{\mathcal{Y}_{12}-\mathcal{Y}_{34}}\right),\\
			v_{Q_2}=&\frac{\beta_v}{2\pi} \log \left(\frac{-\sqrt{\tilde{\mathcal{Y}}_{12} \tilde{\mathcal{Y}}_{13} \tilde{\mathcal{Y}}_{24}\tilde{\mathcal{Y}}_{34}}+\tilde{\mathcal{X}}_{14}-\tilde{\mathcal{X}}_{23}}{\tilde{\mathcal{Y}}_{12}-\tilde{\mathcal{Y}}_{34}}\right),
		\end{aligned}
	\end{equation}
	where the functions $\mathcal{X}_{ij},\tilde{\mathcal{X}}_{ij},\mathcal{Y}_{ij},\tilde{\mathcal{Y}}_{ij}$ take the following form,
	\begin{equation}
		\begin{aligned}
				\mathcal{X}_{ij}=&e^{\frac{2 \pi  (u_i+u_j)}{\beta_u}},\quad 	\mathcal{Y}_{ij}=e^{\frac{2 \pi  u_i}{\beta_u}}-e^{\frac{2 \pi  u_j}{\beta_u}},\\
				\tilde{\mathcal{X}}_{ij}=&e^{\frac{2 \pi  (v_i+v_j)}{\beta_v}},\quad 	\tilde{\mathcal{Y}}_{ij}=e^{\frac{2 \pi  v_i}{\beta_v}}-e^{\frac{2 \pi  v_j}{\beta_v}}.
		\end{aligned}
	\end{equation}
 In the $(U,V)$ coordinate system, the coordinates of $Q_1$ and $Q_2$ can be obtained accordingly by applying the field-dependent coordinate transformations \eqref{coordinatetrans} and the replacements \eqref{nodtem} and \eqref{TMGparameters}. Substituting the coordinates of $Q_1$ and $Q_2$ in the $(U,V)$ coordinate system into \eqref{twopartsBPE}, and subsequently using the ALC proposal \eqref{ALC proposal}, it is straightforward to demonstrate that the BPE exactly aligns with the EWCS with the anomalous correction \eqref{deEWCSnon}.

\section{Summary and Discussions}\label{sec:summary}

	In this work, we developed a holographic framework to study the $\TTbar$ deformation of two-dimensional conformal field theories with gravitational anomalies, using topological massive gravity (TMG) as the dual bulk description. We focused on a regime of small deformation parameter $\mu$, and unequal central charges ($c^{}_L \neq c^{}_R$), characteristic of parity-violating theories. By implementing mixed boundary conditions for the bulk metric as described in \cite{Guica:2019nzm}, we constructed the corresponding deformed BTZ geometry in TMG and obtained the resulting stress tensor via holographic renormalization \cite{Skenderis:1999nb,Balasubramanian:1999re}. This approach enabled us to derive the deformed energy spectrum and verify its consistency with field-theoretic expectations.
	
	Our results confirm that the deformed energy spectrum retains the universal square-root structure found in parity-symmetric $\TTbar$-deformed CFTs \cite{Smirnov:2016lqw,McGough:2016lol,Kraus:2018xrn,Guica:2019nzm}. Moreover, we performed a perturbative analysis of entanglement entropy and reflected entropy from the boundary field theory and found exact agreement with bulk computations using spinning particle worldlines at the first order in the $\mu$ expansion. Notably, we found that the entanglement entropy remains real only if the deformation parameter $\mu$ satisfies a generalized Hagedorn-like bound, now sensitive to the asymmetry between $c^{}_L$ and $c^{}_R$. Furthermore, we computed the BPE by using the non-perturbative holographic entanglement entropy for single intervals and showed it exactly reproduces the EWCS with CS correction. This demonstrates that the fine correspondence between BPE and EWCS persists despite gravitational anomalies and $\TTbar$ deformation.
	
	These agreements provide strong evidence that the holographic dictionary for mixed boundary conditions remains valid even in the presence of a gravitational Chern–Simons term. Our work extends the literature on irrelevant deformations and suggests that solvability and consistency of the $\TTbar$ flow are preserved under gravitational anomalies. The analysis also highlights the efficacy of using spinning probes in the bulk to accessing parity-violating entanglement data, enriching our understanding of quantum information in anomalous systems.
		
		\vspace{1em}
	
	Our investigations open several promising directions for future research:
	\begin{itemize}
		\item \textbf{Beyond leading order:} We have focused primarily on the leading-order corrections in $\mu$ in our field-theoretic analysis. It would be valuable to study higher-order effects, extending the techniques introduced in \cite{Chen:2018eqk,Jeong:2019ylz,Basu:2024enr}.
		
		\item \textbf{Dynamical and time-dependent backgrounds:} Exploring $\TTbar$-deformed anomalous CFTs in non-equilibrium setups, such as quenches or expanding backgrounds, may reveal new features of entropy production, chaos, and thermalization in the presence of anomalies.
		
		\item \textbf{The inner RT Surface in $\TTbar$-deformed holography:} Recently, the authors in \cite{Wen:2024muv} proposed another geometric picture of the holographic entanglement entropy with gravitational anomalies and time-like entanglement entropy, namely the inner RT surface (IRT surface), which is the pre-image of the inner horizon in the Rindler space \cite{Casini:2011kv}. It is interesting to study the IRT surface in the $\TTbar$-deformed holography.
		
		\item \textbf{Chern-Simons formulation and Wilson line observables:} It would be particularly interesting to explore the Chern-Simons formulation of $\TTbar$-deformed AdS$_3$ gravity in the presence of topological massive gravity (TMG), following the approaches developed in \cite{Llabres:2019jtx, He:2020hhm}. In this context, a natural direction is to generalize the results of \cite{He:2023xnb} by incorporating the factorized Wilson line prescription introduced in \cite{Castro:2014tta}.
		
		\item \textbf{Relations to chaos and pole skipping:} Given the recent demonstration that entanglement wedge, OTOCs, and pole skipping all probe the same soft modes in holography \cite{Chua:2025vig}, it would be interesting to test whether this equivalence persists in the presence of gravitational anomalies and $\TTbar$ deformation. This would deepen our understanding of chaos in anomalous field theories.
		
		\item \textbf{Extensions to other deformations:} Analogous studies of $J\bar{T}$, $T\bar{J}$, or other warped deformations \cite{Guica:2017lia,Chakraborty:2018vja,Bzowski:2018pcy,Guica:2019vnb,Kruthoff:2020hsi} in anomalous CFTs would broaden the scope of this program and could reveal new solvable sectors with rich parity-violating dynamics.
		
		\item \textbf{Black hole interiors and extended $\TTbar$ flows:}  
		Recently the authors in \cite{AliAhmad:2025kki} proposed an extended $\TTbar$ deformation that allows pushing the holographic cutoff surface inside the horizon, enabling a controlled probe of the black hole interior without encountering complex energy eigenvalues. It would be interesting to explore whether a similar extension can be formulated in the presence of gravitational anomalies, potentially offering new insights into interior reconstruction in anomalous holography.
		
	\end{itemize}
	
	\vspace{1em}
	
	\noindent In conclusion, this work offers a self-consistent holographic framework for $\TTbar$-deformed anomalous CFTs and demonstrates the compatibility of irrelevant deformations with parity-violating holography. Our results may shed light on future studies of quantum information, entanglement, and non-local observables in a broad class of anomalous but solvable quantum field theories.

	\section*{Acknowledgment}
	The authors are supported by the NSFC Grant No. 12447108 and the Shing-Tung Yau Center of Southeast University. We thank Yunfeng Jiang and Miao He for reading the manuscript and very helpful suggestions.

	\appendix

	\section{Fefferman-Graham expansion and MBC prescription}\label{Sec F-G gauge}
	
	In three dimensions, the most general asymptotic AdS solutions can be written in the radial gauge through the Fefferman-Graham expansion,
	\begin{equation}
		ds^2=\frac{d\rho^2}{4\rho^2}+g_{\alpha\beta}\left(\rho,x^\alpha\right)dx^\alpha dx^\beta,\quad g_{\alpha\beta}\left(\rho,x^\alpha\right)=\frac{g^{\left(0\right)}_{\alpha\beta}\left(x^\alpha\right)}{\rho}+g_{\alpha\beta}^{\left(2\right)}\left(x^\alpha\right)+\rho g_{\alpha\beta}^{\left(4\right)}\left(x^\alpha\right).
	\end{equation} 
	The expansion coefficients $g^{\left(4\right)}_{\alpha\beta}$ and $g^{\left(2\right)}_{\alpha\beta}$ can be algebraically determined from  $g^{\left(0\right)}_{\alpha\beta}$ by the Einstein equations \cite{Skenderis:1999nb},
	\begin{equation}
		g_{\alpha \beta}^{\left(4\right)}=\frac{1}{4}g_{\alpha\gamma}^{\left(2\right)}g^{\left(0\right)^{\gamma\delta}}g^{\left(2\right)}_{\delta\beta}
	\end{equation}
	and
	\begin{equation}
		\text{Tr}\left[\left(g^{\left(0\right)}\right)^{-1}g^{\left(2\right)}\right]=-\frac{1}{2}R\left[g^{\left(0\right)}\right],\quad \nabla^{\left(0\right)}_\alpha g^{\left(2\right)\alpha\beta}=\nabla_\beta g^{\left(2\right)\alpha}_{\alpha}.
	\end{equation}
	Upon performing the holographic renormalization, the coefficient $g^{\left(2\right)}$ turns out to be proportional to the expectation value of the stress tensor of the boundary CFT \cite{deHaro:2000vlm,Balasubramanian:1999re},
	\begin{equation}
		g_{\alpha\beta}^{\left(2\right)}=8\pi G\left(T_{\alpha\beta}-g^{\left(0\right)\alpha\beta}T^{\gamma}{}_{\gamma}\right).
	\end{equation}
	From \eqref{flowsolutions}, one may now obtain the non-linear mixed boundary conditions (involving both $g^{(0)}$ and $g^{(2)}$ as opposed to the usual Dirichlet boundary conditions) for the $\TTbar$-deformed CFT$_2$ \cite{Guica:2019nzm}
	\begin{align}
		\gamma^{[\mu]}_{\alpha\beta}=g^{(0)}_{\alpha\beta}-\frac{\mu}{4\pi G}g^{(2)}_{\alpha\beta}+\frac{1}{4}\left(\frac{\mu}{4\pi G}\right)^2g_{\alpha\gamma}^{\left(2\right)}g^{\left(0\right)^{\gamma\delta}}g^{\left(2\right)}_{\delta\beta}\equiv\rho_c\,g_{\alpha\beta}(\rho_c)\,,
	\end{align}
	with $\rho_c=-\frac{\mu}{4\pi G}$. Incidentally, $\gamma_{\alpha\beta}^{[\mu]}$ could be identified with the induced metric on the $\rho=\rho_c$ slice, in some sense conforming to the cut-off prescription \cite{McGough:2016lol}.
	
	\section{Alternative way of relating $\CL_\mu$ with $\CL_0$} \label{appendix relations}
	
	In this appendix, following \cite{Guica:2019nzm}, we provide an alternative way of relating the deformed charges $(\mathcal{L}_\mu, \bar{\mathcal{L}}_\mu)$ with the undeformed ones. Namely, we find a coordinate transformation that brings the deformed metric \eqref{deformed geometry} into the standard BTZ form while keeping the periodicity of the $\phi$ coordinate unchanged. Such coordinate transformations cannot affect the horizon area and the energy density. Therefore, one is only allowed to rescale $\rho$ and $T$ and shift $\phi$ through a multiple of $T$ as follows\footnote{The rescaling of coordinates can be easily achieved by choosing an ansatz that requires the metric components $g_{\tilde{U}\tilde{U}}$ and $g_{\tilde{V}\tilde{V}}$ to be independent of the radial coordinate $\tilde{\rho}$, together with $g^{(2)}_{\tilde{U}\tilde{V}} = 1/\tilde{\rho}$.}
	\begin{equation}
		\begin{aligned}
			\phi=&\:\tilde{\phi}-2\mu\frac{\kappa\CL_\mu-\bar\kappa\bar\CL_\mu}{1-4\mu^2\kappa\bar\kappa\CL_\mu\bar\CL_\mu}\tilde{T},\\
			\rho=&\:\frac{(1+2\mu\kappa\CL_\mu)(1+2\mu\bar\kappa\bar\CL_\mu)}{(1-4\mu^2\kappa\bar\kappa\CL_\mu\bar\CL_\mu)^2}\tilde{\rho},\\
			T=&\:\frac{(1+2\mu\kappa\CL_\mu)(1+2\mu\bar\kappa\bar\CL_\mu)}{(1-4\mu^2\kappa\bar\kappa\CL_\mu\bar\CL_\mu)}\tilde{T}\label{alternative-map}
		\end{aligned}
	\end{equation}
	In terms of the new null coordinates $\tilde{U},\tilde{V}=\tilde{\phi}\pm\tilde{T}$ and the radial coordinate $\tilde{\rho}$, the deformed metric \eqref{deformed geometry} now takes the form 
	\begin{align}
		\d s^2=\frac{d \tilde{\rho}^2}{4\tilde{\rho}^2}+\frac{d \tilde{U}d\tilde{V}}{\tilde{\rho}}&+\frac{\mathcal{L}_\mu\left(1+2\mu\bar{\kappa}\bar{\mathcal{L}}_\mu\right)^2}{\left(1-4\mu^2\kappa\bar{\kappa}\mathcal{L}_\mu\bar{\mathcal{L}}_\mu\right)^2}d\tilde{U}^2+\frac{\bar{\mathcal{L}}_\mu\left(1+2\mu\kappa\mathcal{L}_\mu\right)^2}{\left(1-4\mu^2\kappa\bar{\kappa}\mathcal{L}_\mu\bar{\mathcal{L}}_\mu\right)^2}d\tilde{V}^2\notag\\
		&+\CL_\mu\bar\CL_\mu\frac{(1+2\mu\kappa\CL_\mu)^2(1+2\mu\bar\kappa\bar\CL_\mu)^2}{(1-4\mu^2\kappa\bar\kappa\CL_\mu\bar\CL_\mu)^4}\tilde{\rho}\,d\tilde{U}d\tilde{V}\,.\label{alternative-bulk-metric}
	\end{align}
	Equating the coefficients of $\d\tilde{U}^2$ and $\d\tilde{V}^2$ with $\CL_0$ and $\bar\CL_0$ respectively, we find precisely the relations \eqref{parelaTMG}. 
	
	
	\section{On induced metric at the asymptotic boundary}\label{Sec induced metric}
	From \eqref{alternative-bulk-metric}, we may easily find the induced metric on the asymptotic boundary at $\tilde{\rho}=\tilde{\rho}_\infty$ as 
	\begin{align}
		\d s^2_\textrm{bdy}=\frac{1}{\tilde{\rho}_\infty}\d\tilde{U}\,\d\tilde{V}=\frac{1}{\tilde{\rho}_\infty}\left(-\d\tilde{T}^2+\d\tilde{\phi}^2\right)
	\end{align}
	From \eqref{alternative-map}, we may rewrite the above metric in terms of the boundary conformal time as
	\begin{align}
		\d s^2_\textrm{bdy}=\frac{-\d T^2+\d\hat{\phi}^2}{\rho_\infty(1+2\mu\kappa\CL_\mu)(1+2\mu\bar\kappa\bar\CL_\mu)}:=\Omega_\omega^{-1}\Omega_{\bar\omega}^{-1}\frac{\d\omega\,\d\bar\omega}{\rho_\infty}
	\end{align}
	where we have switched to the Euclidean signature through the Wick rotation $\tau=i \,T$ and defined the complex coordinate $\omega=\hat\phi+i\tau$, where the conformal spatial coordinate is given as
	\begin{align}
		\hat{\phi}=\frac{(1+2\mu\kappa\CL_\mu)(1+2\mu\bar\kappa\bar\CL_\mu)}{(1-4\mu^2\kappa\bar\kappa\CL_\mu\bar\CL_\mu)}{\phi}+2\mu\frac{\kappa\CL_\mu-\bar\kappa\bar\CL_\mu}{1-4\mu^2\kappa\bar\kappa\CL_\mu\bar\CL_\mu}T\,.
	\end{align}
	Notably, in these coordinates, the boundary metric is conformal to a thermal cylinder, with constant conformal factors for the holomorphic and anti-holomorphic sectors
	\begin{align}
		\Omega_\omega=1+2\mu\kappa\CL_\mu~~,~~\Omega_{\bar\omega}=1+2\mu\bar\kappa\bar\CL_\mu
	\end{align}
	In particular, the thermal identifications in these coordinates are given by
	\begin{align}
		(\omega,\bar{\omega})\sim(\omega+\beta_\omega,\bar{\omega}+\beta_{\bar\omega})
	\end{align}
	where the thermal periodicities are given by
	\begin{align}
		\beta_\omega=\frac{\pi}{\sqrt{\CL_\mu}}+2\pi\mu\kappa\sqrt{\CL_\mu}~~,~~\beta_{\bar\omega}=\frac{\pi}{\sqrt{\bar\CL_\mu}}+2\pi\mu\bar\kappa\sqrt{\bar\CL_\mu}
	\end{align}
	\subsection{Entanglement entropy}
	Now we consider the computation of the entanglement entropy of a single interval $A=[(\phi_1,T_1),(\phi_2,T_2)]$ in the deformed anomalous CFT$_2$. Utilizing the replica trick \cite{Calabrese:2004eu,Calabrese:2009qy}, the R\'enyi entropy can be computed through the two-point correlation function of twist operators inserted at the endpoints of the interval as follows 
	\begin{align}
		S_n(\rho_A)=\frac{1}{1-n}\log\left[\Omega_\omega^{h_{\sigma_n}}\Omega_{\bar\omega}^{\bar h_{\sigma_n}}\langle\sigma_n(\omega_1, \bar{\omega}_1)\bar{\sigma}_n(\omega_2, \bar{\omega}_2)\rangle_\CM\right]
	\end{align}
	We may obtain the correlation function of twist operators by transforming to the complex plane through the following conformal map:
	\begin{align}
		\omega\to z=e^{\frac{2\pi \omega}{\beta_\omega}}~~,~~\bar \omega\to \bar z=e^{\frac{2\pi \bar \omega}{\beta_{\bar\omega}}}\,.
	\end{align}
	Now utilizing the conformal dimensions of the twist operators from \eqref{twist-dimensions} and accounting for the correct conformal factors, we obtain the entanglement entropy as follows
	\begin{align}
		S_A=&\frac{c^{}_L}{6}\log\left[\frac{\beta_\omega}{\pi\Omega_\omega\epsilon}\sinh\left(\frac{\pi\Delta \omega}{\beta_\omega}\right)\right]+\frac{c^{}_R}{6}\log\left[\frac{\beta_{\bar\omega}}{\pi\Omega_{\bar\omega}\epsilon}\sinh\left(\frac{\pi\Delta \bar\omega}{\beta_{\bar\omega}}\right)\right]\notag\\
		=&\frac{c^{}_L}{6}\log\left[\frac{1}{\sqrt{\CL_\mu}\epsilon}\sinh\left(\frac{\sqrt{\CL_\mu}\left(\Delta U+2\mu\bar\kappa\Delta V\right)}{1-4\mu^2\kappa\bar\kappa\CL_\mu\bar\CL_\mu}\right)\right]\notag\\
		&+\frac{c^{}_R}{6}\log\left[\frac{1}{\sqrt{\bar\CL_\mu}\epsilon}\sinh\left(\frac{\sqrt{\bar\CL_\mu}\left(\Delta V+2\mu\kappa\Delta U\right)}{1-4\mu^2\kappa\bar\kappa\CL_\mu\bar\CL_\mu}\right)\right],\label{EE}
	\end{align}
	where in the last equality, we have switched back to the Lorentzian signature. The above result clearly conforms to the bulk computations \eqref{HEETTbar}. The reflected entropy for adjacent or non-adjacent intervals may also be computed in a similar fashion.
	
	\subsection{Strong sub-additivity}\label{Sec stradd}
	
	The strong sub-additivity for the entanglement entropy is given by
	\begin{align}
		S_A+S_B-S_{A\cup B}-S_{A\cap B}\geq 0\,.
	\end{align}
	In the limit where $A\to B$ and $|A|=|B|=L$, this inequality can be expressed in the infinitesimal form \cite{Casini:2012ei}
	\begin{align}
		L^2S^{\prime\prime}(L)\leq 0\,.
	\end{align}
	A stronger criterion for proving locality, compared to the strong sub-additivity, is the boosted strong sub-additivity, which can be derived by applying a boost transformation to subsystems $A$ and $B$. The infinitesimal form of the boosted strong sub-additivity
	inequality is expressed as \cite{Casini:2012ei,Lewkowycz:2019xse}
	\begin{align}
		L^2S^{\prime\prime}(L)+LS^\prime(L)\leq 0\,.
	\end{align}
	For $\Delta U=\Delta V=L$, we may easily find from \eqref{EE}
	\begin{align}
		S''(L)=-&\frac{c^{}_L\bar\CL_\mu}{6}\left(\frac{1+2\tilde\mu \kappa\CL_\mu}{1-4\mu^2\kappa\bar\kappa\CL_\mu\bar\CL_\mu}\right)^2\textrm{csch}^2\left(\frac{\sqrt{\CL_\mu}\left(1+2\mu\bar\kappa\right)L}{1-4\mu^2\kappa\bar\kappa\CL_\mu\bar\CL_\mu}\right)\notag\\&-\frac{c^{}_R\CL_\mu}{6}\left(\frac{1+2\tilde\mu \bar\kappa\bar\CL_\mu}{1-4\mu^2\kappa\bar\kappa\CL_\mu\bar\CL_\mu}\right)^2\textrm{csch}^2\left(\frac{\sqrt{\bar\CL_\mu}\left(1+2\mu\kappa\right)L}{1-4\mu^2\kappa\bar\kappa\CL_\mu\bar\CL_\mu}\right)<0\,,
	\end{align}
	and conclude that the strong sub-additivity is satisfied in TMG with $\TTbar$-deformation for any signature of the deformation parameter and TMG coupling. On the other hand, we find that the boosted strong sub-additivity is always violated for any signature of the deformation parameter, conforming to the non-local nature of the deformed theory.



	\bibliographystyle{JHEP}
	\bibliography{bib}
\end{document}